\documentclass[12pt]{article}
\usepackage{epsfig}
\usepackage[hang,bf,small]{caption}
\DeclareGraphicsExtensions{.eps.gz,.eps,.ps,.ps.gz}
\parskip=\itemsep               

\newcommand{\beq}{\begin{equation}}
\newcommand{\eeq}{\end{equation}}
\newcommand{\beqar}[1]{\begin{eqnarray}\label{#1}}
\newcommand{\eeqar}{\end{eqnarray}}

\newcommand{\as}{\alpha_S}

\def\eq#1{{Eq.~(\ref{#1})}}
\def\fig#1{{Fig.~(\ref{#1})}}
\def\npb#1#2#3{    {\it Nucl. Phys. }{\bf B#1} (19#2) #3}
\def\plb#1#2#3{    {\it Phys. Lett. }{\bf B#1} (19#2) #3}
\def\prd#1#2#3{    {\it Phys. Rev. }{\bf D#1} (19#2) #3}

\def\zpc#1#2#3{    {\it Z. Phys. }{\bf C#1} (19#2) #3}

\relax
\begin{document}
\title{
{\Large \bf   Towards a New Global QCD Analysis: }\\\
{ \Large \bf   Low $x$ DIS Data from Non-Linear Evolution}}
\author{
{\large  ~E. ~Gotsman, \thanks{e-mail:
gotsman@post.tau.ac.il}~~$\mathbf{{}^{a)}}$ E.~Levin,\thanks{e-mail:
leving@post.tau.ac.il;\,\,\,levin@mail.desy.de}~~$\mathbf{{}^{a),\,b)}}$
M. ~Lublinsky,\thanks{e-mail:
mal@post.tau.ac.il;\,\,\,mal@tx.technion.ac.il}~~$\mathbf{{}^{a)}}$
~\,and\,~ ~U. ~Maor\thanks{e-mail:
maor@post.tau.ac.il}~~$\mathbf{{}^{a)}}$
}\\[4.5ex]
{\it ${}^{a)}$ HEP Department}\\
{\it  School of Physics and Astronomy}\\
{\it Raymond and Beverly Sackler Faculty of Exact Science}\\
{\it Tel Aviv University, Tel Aviv, 69978, ISRAEL}\\[2.5ex]
{\it ${}^{b)}$ DESY Theory Group}\\
{\it 22603, Hamburg, GERMANY}\\[4.5ex]
}

\date{\today}
\maketitle
\thispagestyle{empty}

\begin{abstract}
A new approach to global QCD analysis is developed. The main
ingredients are two QCD-based evolution equations. The first one
is the Balitsky-Kovchegov nonlinear equation, which sums higher
twists while preserving unitarity. The second equation is linear
and it is responsible for the correct short distance behavior of
the theory, namely it includes the DGLAP kernel. Our approach
allows extrapolation of the parton distributions to very high
energies available at the LHC as well as very low photon
virtualities, $Q^2\ll 1\, {\rm GeV^2}$.

All existing low $x$ data on the $F_2$ structure function is
reproduced using one fitting parameter. The resulting
$\chi^2/df=1$.

Analyzing the parameter $\lambda\equiv \partial\ln
F_2/\partial(\ln 1/x)$ at very low $x$ and  $Q^2$ well below $1\,
{\rm GeV^2}$ we find $\lambda \simeq 0.08 - 0.1$. A result which agrees
with the "soft pomeron" intercept without involving soft physics.

 \end{abstract}
\thispagestyle{empty}
\begin{flushright}
\vspace{-20.5cm}
TAUP-2713-02\\
DESY 02-133\\
\today
\end{flushright}
\newpage
\setcounter{page}{1}

\section{Introduction}
\setcounter{equation}{0}

In the paper \cite{LGLM} a new approach to DIS was proposed. In
the present paper we review and further develop the ideas
introduced in Ref. \cite{LGLM}. Our main result is that all
existing low $x$ data on the $F_2$ structure function can be
described by nonlinear QCD evolution.

The standard perturbative QCD approach to deep inelastic
scattering (DIS) is based on the DGLAP evolution equation
\cite{DGLAP} which provides the leading twist parton
distributions. The main underlying assumption is that the high
twist contributions are negligibly small if the evolution starts
at sufficiently high photon virtualities $Q_0^2  \approx 2 - 4\,
(\rm GeV^2)$.

The approach based on the DGALP equation suffers from three
principal problems.

\begin{itemize}
\item The DGLAP evolution predicts a steep growth  of parton
distributions in the region of low $x$ which would eventually
contradict the unitarity constraints \cite{GLR}. Hence, we can
expect large unitarity corrections to the DGLAP evolution
equation, in the region of very low $x$.

\item The second problem is the  general nature for any operator
product expansion which is an asymptotic series. In application to
DIS this means that the errors associated with the leading twist
approximation are not small. They are of the order of the next to
leading order twist contribution which  grows very fast at low
$x$. In fact, it can be shown that high twist contributions grow
with decreasing $x$ faster than the leading twist \cite{HT}.
Hence, we cannot conclude that the higher twist contributions are
small in the whole kinematic region, even if they are small for
the initial value of $Q^2 = Q^2_0$. The estimates of Ref.
\cite{HTM} show that all available parameterizations of the
solutions to the DGLAP evolution equation lead to substantial
higher twist contributions.

\item The last problem is in the total nonability of the DGLAP
equation to describe physics of low photon virtuality $Q^2 \le
1\,(\rm GeV^2)$. For these kinematics one needs to use Regge
phenomenology or other phenomenological models.
\end{itemize}
It is important to add that NLO DGLAP, though it improves the fits
to presently available data, does not solve any of the above
principal difficulties. Consequently, we are lead to the
conclusion that DGLAP is insufficient to describe all kinematical
phase space. For small values of $x$ and/or $Q^2$ there is need
for a new QCD-based idea.

In this paper,  we develop such an idea which allows us to
extrapolate parton distributions to very low $x$ (high energies).
An extrapolation of the available parton distribution to the
region of lower $x$ is a practical problem for the LHC energies.
We need to know the parton distribution both for estimates of the
background of all interesting processes at the LHC, such as Higgs
production, and for the calculation of the cross sections of the
rare processes which are likely to be measured at the LHC.

 Our method was originally proposed in Ref. \cite{LGLM}. It
 consists of two steps. As a first step, a nonlinear evolution
 equation,  which takes into account the most significant higher twist
 contributions, is solved. This equation (\ref{EQ}) specifies  a high energy (low $x$)
 behavior of the parton densities. The  solution obtained, below denoted as $\tilde N$,
takes into account collective phenomena of high parton density QCD
and respects  unitarity constraints. Moreover, $\tilde N$ can be
found for large transverse distances, and so it also provides a
possibility to describe data of low photon virtuality.

The parton distributions which we obtain are then amended by
adding  to the solution of the nonlinear equation $\tilde{N}$, a
correcting function $\Delta N $, which is aimed at correctly
incorporating the short distance behavior of the theory, namely
the DGLAP kernel.

For $\Delta N$ we propose  a linear DGLAP-type evolution equation.
The function $\Delta N$ is considered  to be a small correction to
$\tilde N$ concentrated in the region of  moderate $x$.
Consequently, this function should be free of all difficulties
inherent in the usual solutions of the DGLAP equation.

The philosophy of our approach is quite similar to the one
recently presented by Ref. \cite{BGH}. That paper is a development
of the Golec-Biernat - Wusthoff (GBW) model which in addition to
the original model \cite{GW} is improved by  DGLAP evolution. We
can trace a certain  analogy between our function $\tilde N$ and
the original saturation model. Both functions play the very same
role: they take into account the gluon saturation effects in
unitarity preserving way, and describe physics of large distances.
However, contrary to the saturation model  the function $\tilde N$
is derived from QCD. The DGLAP improvement both in our approach
and in Ref. \cite{BGH} is aimed at correctly incorporating the
short distance dynamics  however, the technical realizations are
quite different.

The paper is organized as follows. In the next section we review
our approach and  write the BK  nonlinear equation for $\tilde{N}$ and
a linear equation for  $\Delta N$. Section 3 presents some
analytical estimates of the corrections induced by the DGLAP
kernel. Section 4 is devoted to technical details of numerical
solutions of the  equations. The following Section (5) presents
our fit to the experimental data on $F_2$ structure function and
its logarithmic derivatives. Some predictions for THERA and LHC are given.
Section 6 presents a general  discussion which emphasizes some
 weak points of our approach. In the concluding
Section (7)  we summarize our results and mention our plans for
future work.

\section{Reviewing a new  approach to DIS}

The DGLAP equation describes the gluon radiation which leads to a strong
 increase in  the number of partons. However, when the parton density becomes large,
annihilation processes become important and they   suppress the
gluon radiation  taming the rapid growth  of the parton density
\cite{GLR,MUQI,MV}. A development of new theoretical methods
applicable to physics of high density QCD
\cite{GLR,MUQI,MV,SAT,ELTHEORY,BA,KO,ILM} lead finally to  the
very same nonlinear evolution equation, nowadays credited to
Balitsky and Kovchegov (BK)\footnote{ \eq{EQ} was originally
proposed by Gribov, Levin and Ryskin \cite{GLR} in momentum space
and proven in the double log approximation of perturbative QCD  by
Mueller and Qiu \cite{MUQI}. In the leading $\ln 1/x$
approximation it was derived by Balitsky  in his Wilson Loop
Operator Expansion \cite{BA}. In the form presented in \eq{EQ} it
was obtained by Kovchegov \cite{KO} in the color dipole approach
\cite{MU94} to high energy scattering in QCD. This equation was
also obtained by summation of the BFKL pomeron fan diagrams by
Braun \cite{Braun}
 and in the effective Lagrangian approach for high parton density QCD by Iancu, Leonidov
and McLerran \cite{ILM}. Therefore, it provides a reliable
technique for an extrapolation of the parton distributions to the
region of low $x$.}:

\begin{eqnarray}
\label{EQ}
  \tilde N({\mathbf{x_{01}}},Y;b)&\,=\,&\tilde N({\mathbf{x_{01}}},Y_0;b)\, {\rm exp}\left[-\frac{2
\,C_F\,\as}{\pi} \,\ln\left( \frac{{\mathbf{x^2_{01}}}}{\rho^2}\right)(Y-Y_0)\right ]\,
+\nonumber  \\ & & \frac{C_F\,\as}{\pi^2}\,\int_{Y_0}^Y dy \,  {\rm exp}\left[-\frac{2
\,C_F\,\as}{\pi} \,\ln\left( \frac{{\mathbf{x^2_{01}}}}{\rho^2}\right)(Y-y)\right ]\,\times
\\ \int_{\rho} \,& d^2 {\mathbf{x_{2}}} &
\frac{{\mathbf{x^2_{01}}}}{{\mathbf{x^2_{02}}}\,
{\mathbf{x^2_{12}}}} \nonumber
\left(\,2\,\tilde N({\mathbf{x_{02}}},y;{ \mathbf{ b-
\frac{1}{2}
x_{12}}})-\tilde N({\mathbf{x_{02}}},y;{ \mathbf{ b -
\frac{1}{2}
x_{12}}})\tilde N({\mathbf{x_{12}}},y;{ \mathbf{ b- \frac{1}{2}
x_{02}}})\right) \nonumber
\end{eqnarray}
The equation is derived for $\tilde N(r_{\perp},x; b)$ which
stands for imaginary part of the amplitude of a dipole of size
$r_{\perp}$ elastically scattered at the impact parameter $b$.

In  the equation (\ref{EQ}), the rapidity $Y=-\ln x$ and $Y_0=-\ln
x_0$. The ultraviolet cutoff $\rho$ is needed to regularize the
integral, but it does not appear in physical quantities. In the
large $N_c$ limit (number of colors)   $C_F=N_c/2$ (we set $N_c=3$ in
the numerical computations).

\eq{EQ} has a very simple meaning:  the dipole of size
$\mathbf{x_{10}}$ decays in two dipoles of  sizes
$\mathbf{x_{12}}$ and $\mathbf{x_{02}}$
 with the decay probability
given by  the wave function  $| \Psi|^2
\,=\,\frac{\mathbf{x^2_{01}}}{\mathbf{x^2_{02}}\,\mathbf{x^2_{12}}}$.
 These two dipoles
 then interact with the target. The non-linear term  takes into account
a  simultaneous interaction of two produced dipoles with the
target or, in other words,  the Glauber corrections for
dipole-target  interaction.

The linear part of \eq{EQ} is the LO BFKL equation \cite{BFKL},
which describes the evolution of the multiplicity of the fixed
size color dipoles    with respect to the energy $Y$. The
nonlinear term  corresponds to a dipole splitting into two dipoles
and it sums the high twist  contributions. Note, that the linear
part of \eq{EQ} (the BFKL equation)  also has higher  twist
contributions and vice versa, the main contribution of the
non-linear part is to the  leading twist (see Ref.  \cite{MUQI}
for general arguments and Ref. \cite{HTM} for explicit
calculations).

As has been mentioned,  the master equation (\ref{EQ}) is derived
in the leading $\ln (1/x)$ approximation of perturbative QCD. This
means we consider $\as \ln(1/x) \approx 1$ while $\as \ll 1$ as
well as $\as \ln Q^2 \ll 1$. In other words, the equation sums all
contributions of the order $(\as \ln(1/x))^n$ and neglects
contributions of the orders $\as ( \as \ln(1/x))^n $ and $\as \ln
Q^2 (\as \ln(1/x))^n $. Contributions of the latter will be taken
into account by the function $\Delta N$ to be discussed below.

Eq. (\ref{EQ}) sums all diagrams of the order
$$\left( \,\as^2 (1/x)^{\Delta}\,\right)^n\,\, \mbox{with} \,\,\,\,\Delta
\,\,\propto\,\,\as\,.$$

This means that starting from $\as \ln(1/x)\,\,\approx\,\, \ln
(1/\as)$  corrections due to  rescattering and recombination of
parton become essential (see Ref. \cite{Levin} for details).

The next to leading corrections to the DGLAP or/and BFKL equations
lead to $\Delta = C_1 \as + C_2 \as^2$ which start to be important
only for $\as \ln(1/x) \geq\,1/\as$. Therefore, the
 correct strategy is first to solve  the master
equation taking into account all corrections of the leading order,
and only as a second step consider the next-to-leading order
corrections.

It is well known starting with Bartel's paper in Ref. \cite{HT}
that Eq. (\ref{EQ})  can be proven in the large $N_c$ limit of
QCD. Actually, this equation is a first theoretical realization of
the Veniziano topological expansion \cite{Veneziano}.  For $N_c
\,\,\gg\,\,1$, we assume that $\bar\alpha_s = N_c \as/\pi \approx
1$ while $\as \ll 1$. An interesting feature of the equation is
that it depends  on $\bar \as$ only, and all problems with the
accuracy of the large $N_c$ expansion are really concentrated in
the $N_c$ dependence of the initial distributions (see Ref.
\cite{Levin} for details).

It should be stressed that a correct evolution equation without an
additional assumption on large $N_c$ is known (see Weigert's paper
in Ref. \cite{ELTHEORY}). However, this equation is so complicated
that we are far away from its solution. However, it is worthwhile
mentioning that in the simplified case of double log approach
(both $\as \ln (1/x)$ and $\as \ln Q^2$ are of the order unity
while $\as \,\ll\,1$) the equation for $N_c \approx 1$  was
written and solved in Ref. \cite{Laenen}. This solution shows that
corrections to large $N_c$ approximation are rather small and,
therefore, at the first stage it is reasonable to neglect them.

In order to safely use Eq. (\ref{EQ}) we neede to estimate the
neglected contributions. A first class of such contributions is
the interaction between two parton showers which leads to $(\bar
\as/N^2_c) \ln(1/x)$ corrections which result in a bound on
minimal $x$:
$$
\ln (1/x) \,\,\leq\,\,\frac{N^2_c}{\bar \as}\,.
$$
A second constraint comes from so called enhanced diagrams. It
turns out that they lead to the very same restrictions as the
previous one (see Ref. \cite{Levin} for details).

The above energy limit is not very essential since the unitarity
bound $\tilde N = 1$ is reached at higher values of $x$. Thus the
$1/N_c$ corrections cannot modify this result, but could  slightly
modify the value of the saturation scale.

The total dipole cross section is given by the integration over
the impact parameter: \beq \label{TOTCX} \sigma_{\rm
dipole}(r_{\perp},x) \,\,=\,\,2\,\,\int\,d^2 b\,\,\tilde
N(r_{\perp},x;b). \eeq
The contribution to the deep inelastic
structure function $F_2$ which is due to $\tilde N$ we denote by
$\tilde F_2$ and it is related to the dipole cross section

\beq \label{F2T} \tilde F_2(x,Q^2)\,\,\,=\,\,\frac{Q^2}{4\pi^2
}\,\int\,\,d^2 r_{\perp} \int \,d z\,\, P^{\gamma^*}(Q^2;
r_{\perp}, z) \,\,\sigma_{\rm dipole}(r_{\perp}, x)\,\,. \eeq
 The physical interpretation of \eq{F2T} is transparent. It describes the  two stages of
 DIS \cite{GRIB}. The first stage is
the decay of a virtual photon into a colorless dipole ($ q \bar q
$ -pair). The probability of this decays is given by
$P^{\gamma^*}$.
 The second stage is the interaction of the dipole
with the target ($\sigma_{\rm dipole}$ in \eq{F2T}). This equation
is a simple manifestation of the fact that color dipoles are the
correct degrees of freedom in QCD at high energies \cite{MU94}.
The QED wave functions  of the virtual photon are well known
\cite{MU94,DOF3,WF} (we consider only massless case):

\beq\label{psi}
 P^{\gamma^*}(Q^2; r_{\perp}, z)^2 \,=\,\frac{N_c}{2\,\pi^2}
 \sum_f
 \,Z_f^2\,\left\{(z^2+(1-z)^2)\,a^2\,K_1^2(a\,r_\perp)\,\,+\,\,
 4\,Q^2\,z^2\,(1-z)^2\,K_0^2(a\,r_\perp)\right\}\, ,
 \eeq
with $a^2=Q^2 z (1-z)$.

It can be seen that \eq{EQ} does not depend explicitly on the
target\footnote{This independence is a direct indication that the
equation is
  correct for all targets (hadron and nuclei) in the regime of high parton density.}. All
the dependence on the target comes from the initial condition
specified at some initial value $x_0$. For a target nucleus it was
argued in Ref. \cite{KO} that the initial conditions  should be
  taken in the Glauber form:
\beq \label{ini} \tilde
N(\mathbf{x_{01}},x_0;b)\,=\,N_{GM}(\mathbf{x_{01}},x_0;b)\,, \eeq
with \beq \label{Glauber}
N_{GM}(\mathbf{x_{01}},x;b)\,=\,1\,\,-\,{\rm exp}\left[ -
\frac{\as \pi  \mathbf{x^2_{01}}} {2\,N_c\, R^2}\,x G^{DGLAP}(x,
4/\mathbf{x_{01}^2})\,S(\mathbf{b})\right]\,. \eeq The equation
(\ref{Glauber}) represents the Glauber-Mueller  (GM) formula which
accounts for the  multiple dipole-target interaction in the
eikonal approximation \cite{DOF3,DOF1,DOF2}. The function $S(b)$
is a  dipole  profile function inside the target. The value  of
$x_0$ is chosen within the interval \beq \label{INCON} {\rm exp}(
- \frac{1}{\as} ) \,\leq \,x_0 \, \leq \frac{1}{2 m R}\,\,, \eeq
where $R$ is the radius of the target. In this region the value of
$x_0$ is small enough to use the low $x$ approximation, but the
production of the gluons (color dipoles) is still suppressed as
$\alpha_S \ln (1/x) \,\leq\,\,1$. Consequently, in this region we
have the instantaneous exchange of the classical gluon fields.
Hence, an incoming color dipole interacts separately with each
nucleon in a nucleus (see Mueller and Kovchegov paper in Ref.
\cite{SAT}).

For the hadron, however, there is no proof that \eq{ini} is
correct. Our criteria in this problem (at the moment) is the
correct description of the experimental data. Almost all available
HERA data can be described using \eq{ini} \cite{rep,me},  and we
feel confident setting \eq{ini} as an initial condition for
\eq{EQ}. In our model, the Gaussian ($S(b)=e^{-b^2/R^2}$) form for
the profile function of the hadron is mostly used. The parameter
$R$ is a phenomenological input, while the gluon density
$xG^{DGLAP}$ is a solution of the DGLAP equation. For a hadron
target  \eq{INCON} is still correct, but practically
 $x_0 = 10^{-2}$ is chosen.  This value satisfies \eq{INCON}  for which
much experimental data exist, so one can check the initial
conditions.

Solutions to the BK equation  were studied in asymptotic limits in
Ref. \cite{LT} while several numerical solutions were reported in
Refs.  \cite{Braun,LGLM,Braun2,LL,GMS}.  In Ref. \cite{LGLM} and
in present paper we solve Eq.  (\ref{EQ}) in the coordinate
representation in which the initial conditions are of a very
simple form (see \eq{ini}). The second reason for using the
coordinate representation is the fact that all physical
observables can be expressed in terms of the amplitude for the
dipole-target interaction in the coordinate representation.
Finally, it is also very useful that the long distance
asymptotics is known: $\tilde{N}\,\,\rightarrow\,\,1$ being
$\tilde{N}\,\,\leq\,\,1$ otherwise. This fact provides a natural
control for the numerical procedure.

Unfortunately,  Eq. (\ref{EQ}) is an approximation. It only sums
large $\ln x$ contributions. The situation can be improved at
short distances. The exact $x$ dependence of the kernel at short
distances is known, namely it is the  DGLAP kernel.
 An  attempt to obtain
the elastic amplitude $\tilde N$  based on  elements of  both the
BK and  DGLAP equations was presented in Ref.
\cite{KKM}. The authors of this paper first solve  a generalized
DGLAP-BFKL linear equation \cite{KMS}, and then add to the
solution a nonlinear perturbation of the form presented in Eq.
(\ref{EQ}). This approach actually incorporates the  high twist
contributions
 in the standard way, treating them as corrections
to the leading one.

We suggest a different approach to the problem. First,  all twist
contributions should be summed by solving  Eq. (\ref{EQ}).
Unfortunately, it is  complicated to find a solution for an arbitrary
value of the impact parameter $b$. We simplify the problem by
solving Eq. (\ref{EQ}) without including the $b$-dependence, this
corresponds to case of solving for the initial condition at $b=0$.
At the very end we restore the $b$-dependence by using an ansatz
to be discussed in Section 4. We denote by $\tilde N(r_\perp,x)$
the solution of Eq. (\ref{EQ}) at $b=0$.

\begin{figure}[htbp]
\begin{minipage}{10.0cm}
\epsfig{file=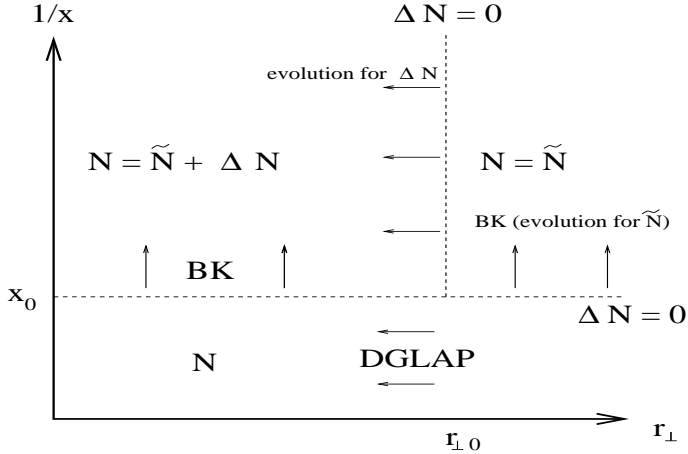,width=90mm, height=60mm}
\end{minipage}
\begin{minipage}{6.0 cm}
\caption{ \it The kinematic map for the solutions.} \label{map}
\end{minipage}
\end{figure}

Secondly, we add to the  solution obtained a correcting function
$\Delta N$, which will account for the DGLAP kernel (Fig.
\ref{map}):

\beq \label{Add} N\,=\,\tilde N\,+\,\Delta N\,\,\,\,\,\,\,\,\,\,\
\,\,for\,\,\, x\,\le\,x_0\,. \eeq

In order to extract the leading $\ln\, (1/r_\perp^2)$ contributions we
define a set of new functions:

$$\tilde n\,\equiv \, \tilde N /(\alpha_s\, r_\perp^2)\,;\,\,
\,\,\,\,\,\,\,\,\, n\,\equiv \,  N /(\alpha_s\,
r_\perp^2)\,;\,\,\,\,\,\,\,\,\, \,\, \, \Delta n\,\equiv \, \Delta
N /(\alpha_s\, r_\perp^2)\,.$$

For the function $n$ we propose the following nonlinear equation
assumed to be valid in the leading large $\ln\, (1/r_\perp^2)$
approximation:

\beq \label{EQN} \frac{\partial n(r_\perp,x)} {\partial\,\ln
(1/r_\perp^2) }\,=\,\frac{C_F\,\alpha_s}{\pi} \int_{x}^1
\,P_{g\rightarrow g}(z)\,\, n(r_\perp, \frac{x}{z})\,dz\,\,-\,\,
\frac{C_F\,\alpha_s^2\,r^2_\perp }{\pi} \int_{x/x_0}^1\frac{d
z}{z} \, n^2(r_\perp, \frac{x}{z})\,. \eeq

Here $ P_{g\rightarrow g}(z)$ stands for the usual gluon splitting
function:

\beq P_{g\rightarrow
g}(z)\,=\,2\,\left[\frac{1-z}{z}\,+\,\frac{z}{(1-z)_+}\,+\,z\,(1-z)\,+\,
\left(\frac{11}{12}-\frac{n_f}{18}\right)\delta(1-z)\right]\,.
\label{split} \eeq

Note that we assume nonlinear effects to be of no importance for
$x>x_0$. Eq. (\ref{EQ}) can be rewritten in the large $\ln
(1/r_\perp^2)$  approximation as:

\beq\label{EQDLA} \frac{\partial \tilde n(r_\perp,x)}
{\partial\,\ln (1/r_\perp^2) }\,=\,\frac{\partial \tilde
n(r_\perp,x_0)} {\partial\,\ln (1/r_\perp^2) } \,+\,
\,\frac{C_F\,\alpha_s}{\pi} \int_{x}^1 \,\frac{dz}{z}\,\,\tilde
n(r_\perp, \frac{x}{z})\,\,-\,\, \frac{C_F\,\alpha_s^2\,r^2_\perp
}{\pi} \int_{x/x_0}^1\frac{d z}{z} \, \tilde n^2(r_\perp,
\frac{x}{z})\,.
 \eeq

Subtracting Eq. (\ref{EQDLA}) from Eq. (\ref{EQN}) and assuming
$\Delta N$ to be small compared to $\tilde N$, we derive the
equation for $\Delta n(r_\perp,x)$:

\begin{eqnarray}
& &\frac{\partial \Delta n (r_\perp, x)}{\partial\, \ln
(1/r_\perp^2) }\,=\,\frac{C_F \as}{\pi} \int_{x/x_0}^1
\,P_{g\rightarrow g}(z)\,\, \Delta n(r_\perp, \frac{x}{z})\,dz
 -  \label{DN} \\ & &\frac{2\,C_F\,\alpha_s}{\pi}
\int_{x/x_0}^1\frac{d z}{z} \,\tilde N(r_\perp, \frac{x}{z})
\Delta n(r_\perp, \frac{x}{z}) + \frac{C_F\,\alpha_s}{\pi}\,
\int_{x/x_0}^1 \left(P_{g\rightarrow g}(z)-\frac{2}{z}\right)
\tilde{n}(r_\perp, \frac{x}{z})\,dz \nonumber \\
& & -\frac{\partial \tilde n(r_\perp,x_0) }{\partial\,\ln
(1/r_\perp^2)}\,\,+\,\, \frac{C_F\,\alpha_s}{\pi}
\int_x^{x/x_0\,-} \,P_{g\rightarrow g}(z)\,\, n(r_\perp,
\frac{x}{z})\,dz\,. \nonumber
\end{eqnarray}

 Equation (\ref{DN}) is a linear equation valid in the
leading $\ln (1/r_\perp^2)$ approximation. The first term on the
right hand side of Eq. (\ref{DN}) is the DGLAP evolution for the
correcting function $\Delta N$, while the second term is the
''nonlinear interaction'' of the solutions. The third term in the
equation represents the correction which is due to the
substitution of the BFKL kernel $1/z$ by the correct DGLAP kernel.

The last term in (\ref{DN}) is the only term which accounts for
the contribution of high $x > x_0$. In that region the function
$n=\pi\,xG^{DGLAP}/(2 N_c R^2)$ being a solution of the DGLAP
equation is not given by a sum of $\tilde n$ and $\Delta n$.
 The sign "-" in the upper integration limit
indicates that in the limit $x \rightarrow x_0$ the
$\delta$-function term of the splitting function must be excluded.

If we set $\tilde N=0$ then Eq. (\ref{DN}) reduces exactly to the
gluonic part of the leading order DGLAP equation. In planned
further development of our approach quark distributions and their
evolution will also be included. At the present stage we take them
into account implicitly in $\alpha_s$ and by setting $n_f=3$ in
$P_{g\rightarrow g}$.

The last two terms in Eq. (\ref{DN}) were omitted in Ref.
\cite{LGLM} as they are not important at low $x$. However they
should be included for correct  computations.

The initial condition $\Delta N(r_{\perp 0},x)=N(r_{\perp
0},x)-\tilde N(r_{\perp 0},x)$ is a phenomenological input at some
initial transverse distance $r_{\perp 0}$ to be specified. However, our
strategy is based on the assumption that $\tilde N$ describes
 the long distance physics correctly. Consequently, in order to
eliminate any discontinuity of the function $N$ we  require
\beq\label{DNINI}
 \Delta N(r_{\perp\, 0},x)=0\,.\eeq
The value of $r_{\perp\, 0}$ is not specified but it is expected
to be of the order $2\, (\rm GeV^{-1})$ corresponding to the naive
relation $Q_0^2=4/r^2_{\perp\, 0}\simeq 1\,(\rm GeV ^2)$.

Since we assume $\tilde N(r_\perp,x_0)$ to describe correctly the
data at $x=x_0$, the continuity across $x=x_0$ supposes that
$\tilde N(r_\perp,x_0)\,=\,N(r_\perp,x_0)$. If we require this
equality then Eq. (\ref{DN}) would respect it provided  initial
condition (\ref{DNINI}) is imposed.

It is important to emphasize that Eq. (\ref{DN}) is free from the
main problems of the DGLAP equation. First of all, the high twist
contributions are summed (at least partially) by Eq. (\ref{EQ}).
Secondly, our method respects unitarity. This is achieved due to
the second term of Eq. (\ref{DN}) and the unitarity preserving
initial condition (\ref{DNINI}).

Finally it is necessary to compute a correction to $F_2$ structure
function due to $\Delta N$. To achieve this goal we need to assign
an impact parameter dependence of $\Delta N$. Similarly to what is
common in DGLAP solutions, the $b$-dependence is assumed to be a
product of $(2\,\pi\,R^2)\Delta N$ times a profile function. After
$b$-integration the latter contributes unity. Then the correction
to $F_2$ reads

\beq\label{DF2} \Delta
F_2(x,Q^2)\,=\,\frac{4\,N_c}{9\,\pi^3}\,\int_{4/Q^2}^{r_{\perp\,0}}\,\frac{d
r_\perp^2}{r_\perp^4}\,\,\Delta N(r_\perp, x)\,(\pi\,R^2)\,. \eeq

The $r_\perp$ integration in (\ref{DF2}) is evaluated in the same
large $\ln 1/r_\perp$ approximation as is valid for Eq.
(\ref{DN}). Note that the coefficient $(\pi\,R^2)$ is cancelled
with the same factor hidden in $\Delta N$.

\section{DGLAP correction - analytical estimates}

In this section we would like to make some comments regarding the
 consistency of our approach. It was argued previously that it is
 necessary to add a
correction term   $\Delta N$  to the solution $\tilde N$ of the
nonlinear equation (\ref{EQ}), where the function $\Delta N$ is a
solution of the evolution equation (\ref{DN}).

Consistency of the approach requires the function  $\Delta N$ to
give vanishing contributions to the dipole cross section at very
small $x$. We also expect this function to decrease as $r_\perp^2$
decreases. Finally, $\Delta N$ is assumed to be a small correction
to the function $\tilde N$. In order to check the above conditions
some asymptotic estimates can be made without explicitly solving
Eq. (\ref{DN}). Indeed, we will show below that Eq. (\ref{DN})
respects all the above mentioned requirements.

\begin{itemize}
\item {\bf Limit 1}: fixed $r_\perp$, $x\rightarrow 0$.

 At very small $x$ and fixed distances the function $\tilde N\simeq 1$.
 Eq. (\ref{DN}) can be simplified:

 \beq \frac{\partial \Delta n (r_\perp, x)}{\partial
\ln (1/r_\perp^2) }\,=\,\frac{C_F \as}{\pi} \int_{x/x_0}^1
\,\left(P_{g\rightarrow g}(z)-\frac{2}{z}\right)\, \left(\tilde n
(r_\perp, \frac{x}{z})\,+\, \Delta n(r_\perp,
\frac{x}{z})\right)\,dz \,. \label{DN1} \eeq

The main observation is that the evolution kernel entering the
equation (\ref{DN1}) is actually negative. Hence the function
$\Delta n$ decreases as $r_\perp$ decreases.

Let us consider a model where the anomalous dimension has the form
consistent with  energy conservation \cite{EKL}

\beq \label{andim} \gamma(\omega)\,= \,\bar \as
\left(\frac{1}{\omega}\,-\,1\right)\,, \eeq

where $\bar \as\equiv \as N_c/\pi$ and the anomalous dimension is
defined by the Mellin transform of the splitting function
$P_{g\rightarrow g}$:

\beq \label{mellin}
\gamma(\omega)\,=\,\frac{\as\,C_F}{\pi}\int_0^1\,dz\,P_{g\rightarrow
g}(z)\,z^\omega. \eeq

Eq. (\ref{DN1}) can be solved using the inverse Mellin transform
\cite{LGLM}. Define $\Delta n(r_\perp,\omega)$ and
$\tilde n(r_\perp,\omega)$ as the inverse Mellin
transforms
$$
\Delta n(r_\perp,\omega)\,\equiv\,\frac{1}{2\,\pi\,
i}\,\int_C\,d\omega\, x^{-\omega}\,\Delta
n(r_\perp,\omega)\,;\,\,\,\,\,\,\,\,\,\,\,\,\,\,\,\, \tilde
n(r_\perp,\omega)\,\equiv\,\frac{1}{2\,\pi\, i}\,\int_C\,d\omega\,
x^{-\omega}\,\tilde n(r_\perp,\omega)\,.
$$
In the momentum representation Eq. (\ref{DN1}) together with the
anomalous dimension (\ref{andim}) is: \beq \label{DNmellin}
\frac{\partial \Delta n(r_\perp,\omega)}{\partial \ln
(1/r_\perp^2)
  }\,=\,-\bar{\alpha}_s\,[\tilde n(r_\perp,\omega)\,+\,\Delta n(r_\perp,\omega)]
\eeq

Eq. (\ref{DNmellin}) can be easily integrated. Applying again the
approximation $\tilde N\simeq 1$ we get
the result for the correcting function $\Delta N$:

\beq \label{solDN} \Delta
N(r_\perp,x)\,\simeq\,-\frac{\bar\as}{1+\bar\as}\tilde
N(r_\perp,x)\,+\, C(x)\,(r_\perp^2)^{1+\bar\as}\,. \eeq

The function $C(x)$ should be determine from the initial condition
$\Delta N(r_{\perp\,0},x)=0$. Consequently

\beq \label{solDN1} \Delta
N(r_\perp,x)\,\simeq\,-\frac{\bar\as}{1+\bar\as}\,\left [\tilde
N(r_\perp,x)\,-\,\tilde N(r_{\perp\,0},x)\,
(r_\perp^2/r^2_{\perp\,0})^{1+\bar\as}\right]\,. \eeq

As expected the function $\Delta N$ is negative and of the order
$O(\as)$ compared to $\tilde N$. As $r_\perp$ decreases, $\Delta
N$ decreases until it reaches a minimum at $r_{\perp\,min}$
determined by the equation

\beq\label{min} \frac{\partial  \tilde N(r_{\perp},x)}{\partial
r_\perp^2}|_{r_\perp=r_{\perp\,min}}\,\simeq\, (1+\bar\as)\,\tilde
N(r_{\perp\,0},x)\,\frac{(r_\perp^2)^{\bar\as}}{(r_{\perp\,0}^2)^{1+\bar\as}},.
\eeq At shorter distances $\Delta N$ tends to $0$ as it should.

Note that at fixed $r_\perp$, $\Delta N$ is finite and non
vanishing at $x\rightarrow 0$.  Yet, this is consistent with the
requirement of a vanishing contribution to the dipole cross
section since the latter implies  integration over the impact
parameter $b$. We will assign different $b$-dependences to the
functions $\tilde N$ and $\Delta N$. After the integration, the
dipole cross section due to $\tilde N$ will grow logarithmically
with $x$, while the contribution of $\Delta N$ will remain finite.

\item {\bf Limit 2}: fixed $x$, $r_\perp\rightarrow 0$.

We would now like  to address a question of the short distance
asymptotics. In this limit, the function $\tilde N$ is given by
the solution of the BFKL equation. Namely,

\beq\label{Nas} \tilde N(r_\perp,\omega)\,\sim\,e^{\,\,\bar\as
\,\ln (1/r_\perp^2)/\omega}\,. \eeq

On the other hand, energy conservation (\ref{andim}) requires
$$
N(r_\perp,\omega)\,\sim\,e^{\,\,\bar\as \,\ln
(1/r_\perp^2)\,(1/\omega-1)}\,.
$$
Hence \beq\label{DNas}
 \Delta N(r_\perp,\omega)\,\simeq\,\tilde
N(r_\perp,\omega)\,(e^{\,\,-\,\bar\as \,\ln
(1/r_\perp^2)}\,-\,1)\,\rightarrow\,-\,\tilde N(r_\perp,\omega).
\eeq

\end{itemize}

Finally we conclude that the function $\Delta N$ is supposed to be
negative. We expect $\Delta N\,\simeq\, \beta \,\tilde N$ with
$|\beta| < 1$ and to be approximately  $x$ independent. At short
distances $\Delta N$ tends to zero as $\tilde N$.

 Therefore, $\Delta N$
turns out to be small in the whole kinematic region. The  analysis
presented above justifies the self consistency of our approach and
paves the way for numerical calculations to be presented in the
next section.

\section{Numerical solution of the equations}

In this section we report on the exact numerical solution of Eq.
(\ref{EQ}) with  initial condition (\ref{ini}) and of Eq.
(\ref{DN}) with  initial condition (\ref{DNINI}).

First of all we wish to discuss several technical details.

\quad $\bullet$ {\bf Kinematic domain}

 The kinematic region where the solutions of (\ref{EQ}) and
 (\ref{DN}) are found, covers  $x$ values from $x=x_0=10^{-2}$, where the initial conditions
are set,   to $x = 10^{-7}$. The maximal transverse distance
$r_\perp$ is taken to be two fermi. The value of the ultraviolet
cutoff $\rho$ is $2\times 10^{-3}\,({\rm GeV^{-1}})$. The
numerical solutions obtained are checked and are independent of
this choice.

\quad $\bullet$ {\bf Coupling constant $\as$}.

Eq. (\ref{EQ}) is derived for constant $\as$. However, the DGLAP
equation and hence \eq{EQN}  have a running coupling constant.
Consequently, the derivation of \eq{DN}  implies the same running
$\as$ in \eq{EQ}. For numerical purposes we take the LO running
$\as$  with $\as=\as (4/r_\perp^2)$ everywhere. At large distances
we freeze $\as$ at the value $\as \simeq 0.5$.

\quad $\bullet$ {\bf Transverse hadron size $R^2$.}

In Ref. \cite{LGLM} the fixed value $R^2=10\,({\rm GeV}^{-2})$ was
taken. This choice corresponds to the value which is obtained from
the ``soft" high energy phenomenology \cite{DL,GLMSOFT}, and is in
agreement with the HERA data on elastic $J/\Psi$ photo-production
\cite{EPP}. As $R^2$ is practically the only fitting parameter at
our disposal  we allow it to vary in order to fit the $F_2$ data.
The optimal fit is achieved at the value $R^2\simeq 3.1\,({\rm
GeV}^{-2})$. This value is too small and requires our
understanding.  The physical meaning of such a small value will
be considered in the Discussion.

\quad $\bullet$ {\bf Gluon density $xG^{DGLAP}$.}

In our approach, the gluon density $xG^{DGLAP}$ appears twice:
first, in initial condition (\ref{ini}) and, second, it accounts
for the region $x > x_0$ in Eq. (\ref{DN}). At this stage, we do
not solve the DGLAP equation for the high $x$ region. Instead, we
rely on the existing parton distributions. Practically for
$xG^{DGLAP}(x\ge x_0,4/r^2_\perp)$ we use the LO CTEQ6
parametrization \cite{CTEQ}.

\quad $\bullet$ {\bf Solution of Eq. (\ref{EQ}).}

\begin{figure}[htbp]
\begin{tabular}{c c c c}
$x=10^{-2}$  & $x=10^{-3}$ & $x=10^{-5}$  & $x=10^{-6}$  \\
\epsfig{file=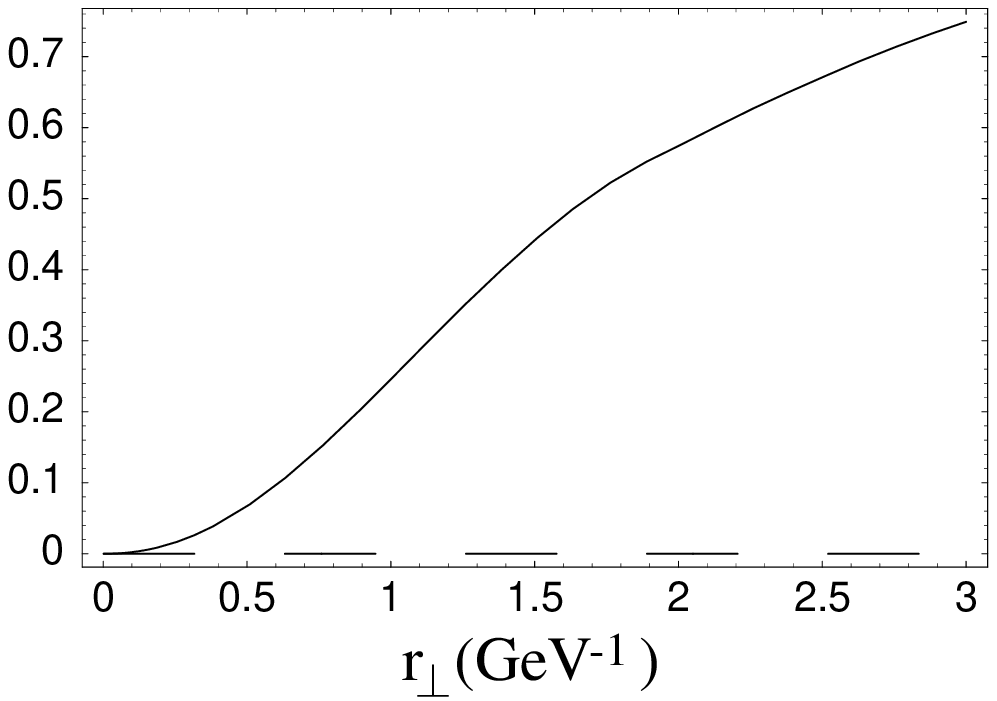,width=38mm, height=34mm}&
\epsfig{file=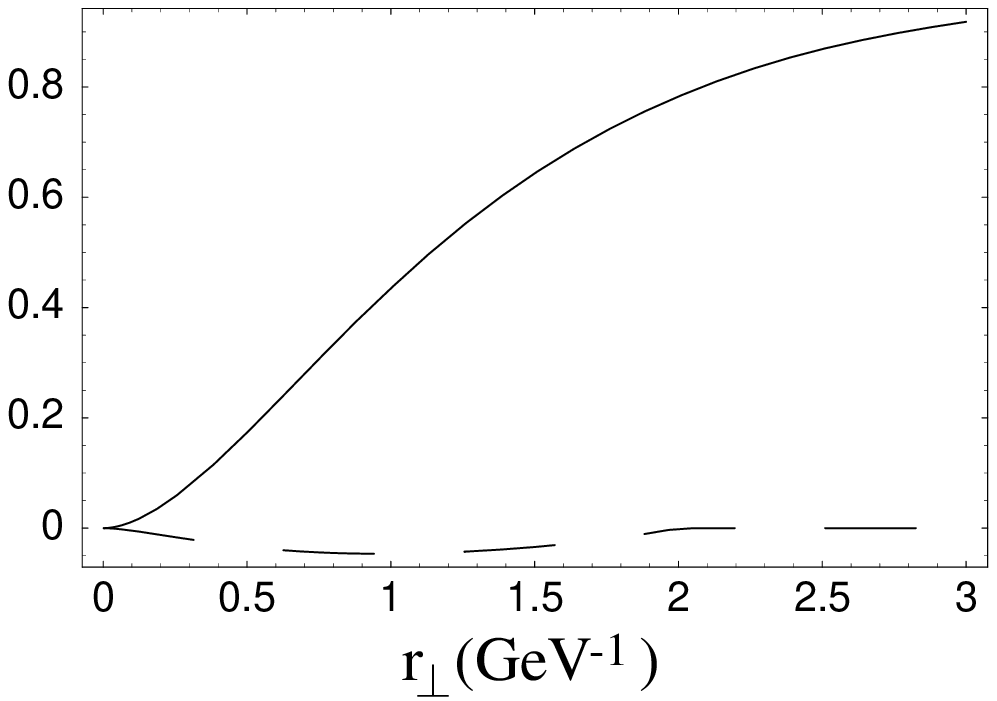,width=38mm, height=34mm}&
\epsfig{file=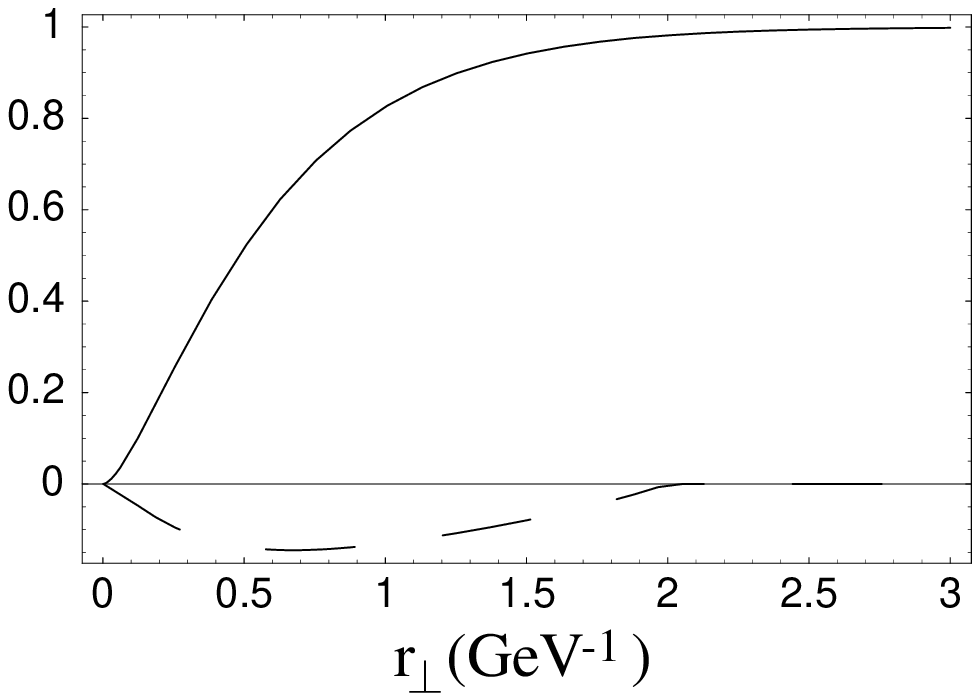,width=38mm, height=34mm}&
 \epsfig{file=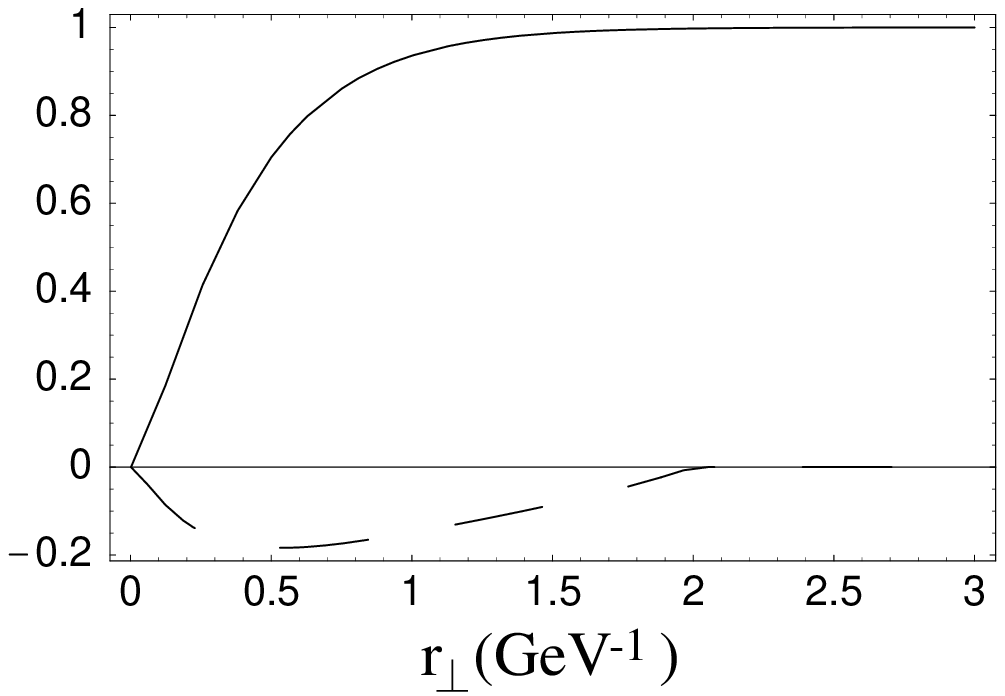,width=38mm, height=34mm}  \\
\end{tabular}
  \caption[]{\it The solution of Eq. (\ref{EQ}) (solid line) and Eq. (\ref{DN}) (dashed line)
   plotted versus $r_\perp$.}
    \label{solution}
\end{figure}

In Ref. \cite{LGLM} Eq. (\ref{EQ}) was solved by the method of
iterations. In the present work we adopt another method which
appears to be more efficient. Namely, we solve Eq. (\ref{EQ}) as
an evolution equation in rapidity with a fixed grid in $r_\perp$
space and a dynamical step in $Y$. The results of the new program
are in total agreement with the old method of Ref. \cite{LGLM}
provided the very same initial input is used. The function $\tilde
N$ is shown in Fig. \ref{solution} (solid curves). At large
distances, $\tilde N$ saturates to unity, which is the unitarity
bound. At short distances, $\tilde N$ tends to zero indicating 
color transparency.

\quad $\bullet$ {\bf Large distances.}

It is of crucial importance for our purposes to correctly
determine $\tilde N$ at large transverse distances. However,
initial conditions (\ref{ini}) are not computable at large
distances. The gluon parametrization appearing in Eq.(\ref{ini})
ends at $r_\perp \simeq 0.5\, fm$. A resolution of the problem was
suggested in Ref. \cite{me1}. The function $\tilde N$ possesses a
property called geometrical scaling. Namely, $\tilde N(r_\perp,x)$
is not a function of two independent variables but rather a
function of single variable $\tau=r_\perp\,Q_s(x)$. Here $Q_s(x)$
stands for the saturation momentum scale.

The solution $\tilde N$ presented in Fig. \ref{solution} is
obtained in two steps. First, initial conditions (\ref{ini}) are
extrapolated to long distances by a constant, which at very long
distances does not approach unity (such a procedure is not
consistent with  scaling that is a purely dynamical property of
the evolution equation). Then a solution is obtained for all $x$.
At sufficiently low $x$ ($x \le 10^{-4}$) the initial conditions
are forgotten and the dynamics is governed by pure evolution. In
this region the geometrical scaling is manifested. As a second
step, we take the solution thus obtained at $x\simeq 10^{-6}$ and
use geometrical scaling in order to rescale this solution up to
$x=x_0$. The resulting curve is now used for a new long distance
extrapolation of initial condition (\ref{ini}). The initial
condition obtained this way  provides a smooth extrapolation of
the Glauber formula to unity at very large distances.

Finally, we note that the  procedure  presented above can be used
for large distance extrapolation of the gluon density at high $x >
x_0$.

\quad $\bullet$ {\bf $b$-dependence of the solution.}

We assume that solution of the equation (\ref{EQ}) preserves the
same $b$-dependence as introduced by the initial conditions
(\ref{ini}):

\beq \label{Nb} \tilde N(r_\perp,x; b)\,=\,
(1\,-\,e^{-\kappa(x,r_\perp)\, S(b)})\,, \eeq where $\kappa$ is
related to the $b=0$ solution

\beq \label{kappa}
\kappa(x,r_\perp)\,=\,-\,\ln(1\,-\,\tilde N(r_\perp,x,b=0)). \eeq

A factorized form of the $b$-dependence was recently advocated
in Ref. \cite{Mac}.
 The ansatz is quite good at moderate
$x$, though it becomes worse at smaller $x$ \cite{LGLM,LL}. The
overall uncertainty of the approximation can be roughly estimated
not to exceed 10\%-20\%.

We now proceed with the evaluation of the dipole cross section
\eq{TOTCX}.
 Having assumed (\ref{Nb}), the dipole cross section has the
form

 \beq \label{sidipol1} \sigma_{\rm
dipole}\,=\,2\,\pi\,R^2\,\left[
\ln(\kappa)\,+\,E_1(\kappa)+\gamma\right]\,. \eeq

In  equation
(\ref{sidipol1}) $\gamma$ denotes  the Euler constant, while
$E_1$ is the exponential integral function. The expression
(\ref{sidipol1}) predicts the $\ln{\kappa}$ growth of the dipole
cross section, which is in agreement with the conclusions
presented in Ref. \cite{LT}.

\quad $\bullet$ {\bf Continuity at $x=x_0$.}

In Section 2 we discussed  the continuity of the function $N$ at
$x=x_0$. One of the ways for its realization is to require $\Delta
N(r_\perp,x_0)=0$ which is fulfilled when $N(r_\perp,x_0)=\tilde
N(r_\perp,x_0)$. However, $\tilde N(r_\perp,x_0)$ is given by GM
formula (\ref{ini}) plus the large distance extrapolation while
$N\sim\as\,r_\perp^2\,xG^{DGLAP}$. Formally they do not coincide
though numerical differences are not significant. Nevertheless, we
decided to force the equality $N(r_\perp,x_0)=\tilde
N(r_\perp,x_0)$. To achieve this, the following changes were
introduced in Eq. (\ref{DN}).

    $$a) \,\,\,\,\, \,\,\,\,\,\,\,\,\,N \,\,\rightarrow\,\, GM\,;
    \,\,\,\,\,\,\,\,\,\,\,\,\,\,\,\,\,\,
    b) \,\,\,\,\, \,\,\,\,\frac{\partial \tilde n(r_\perp,x_0)}{\partial \ln
    (1/r_\perp^2)}\,\,\rightarrow\,\,\frac{C_F\,\as}{\pi}\,\,
    \int_{x_0}^1\,dz\,P_{g\rightarrow g}(z)\,n(r_\perp,x_0/z)\,.$$
The above changes are minor. Practically they affect only the long
distance behavior of the theory for $x\ge x_0$, which is not
significant.

 \quad $\bullet$ {\bf Solution of Eq. (\ref{DN}).}

Having obtained the function $\tilde N$ we can search for the
correction $\Delta N$. Eq. (\ref{DN}) is solved similarly to Eq.
(\ref{EQ}), but with a fixed grid in rapidity and dynamical step
in $r_\perp$. The initial conditions (\ref{DNINI}) are set at
$r_\perp=r_{\perp\,0}$. The parameter $r_{\perp\,0}$ is an
adjusting parameter to be determined from the optimal fit. The
dashed curves in Fig. \ref{solution} show the correcting function
$\Delta N$ corresponding to $r_{\perp\,0}=2\, ({\rm GeV^{-1}})$.

The function $\Delta N$ obtained displays all the qualitative
properties deduced analytically. As expected, starting from zero
at $r_{\perp\,0}$ the function $\Delta N$ decreases until it
reaches a minimum and then it increases to zero again at
asymptotically short distances. The ratio $|\Delta N|/\tilde N$
 increases permanently during the evolution reaching about 70\% at
the edge of our kinematic domain. Below $x\simeq 10^{-3}$ this
ratio  is almost $x$-independent.

\section{Results}

\subsection{Fitting strategy}

Low $x$ $F_2$ data is used to determine the parameters of our
model. The experimental data for $x \le 10^{-2}$ is taken from
ZEUS \cite{ZEUSF2L, ZEUSF2H}, H1 \cite{H1F2}, and E665
\cite{E665F2} experiments. The overall number of points is about
345. We actually use the very same data as Ref. \cite{BGH}.
Statistical and systematic errors are added in quadrature. Whole
data sets are allowed to be shifted within the overall
normalization uncertainity. We use this freedom to shift the low
$Q^2$ ZEUS data down by 2\% and E665 data by 3\%. The H1 data were
shifted up by 3\%.

Our fitting procedure is divided in two steps. First, recall that
in our approach the function $\tilde N$ is supposed to describe
correctly the kinematical region of very low $x$ and large
$r_\perp$. The only fitting parameter for $\tilde N$ is $R^2$. We
vary $R^2$ in order to find an optimal fit for the $F_2$ subset of
data below $Q^2\simeq1\,(\rm GeV^2)$ (about 100 points). The
resulting  $\chi^2/df=1.2$ is achieved for $R^2=3.1\, (GeV^{-2})$.

The function  $\tilde N$ obtained by fitting low $Q^2$ subset of
the data, is not capable of describing all data points. The fit to
all points leads to $\chi^2/df > 3 $, which is not good. The
reason for this mismatch is certainly due to the absence of the
DGLAP kernel in the evolution of $\tilde N$. In order to solve
this problem we switch on the DGLAP correction $\Delta N$ which is
our second step on the way to the optimal fit. To achieve this Eq.
(\ref{DN}) is solved.

For $\Delta N$ the only fitting parameter  is the position
$r_{\perp\,0}$ at which the initial conditions (\ref{DNINI}) are
set. It appears, however, that variation of this position acts as
a fine tuning parameter only. The optimal fit is realized at
$r_{\perp\,0}=2\, (\rm GeV^{-1})$ in total agreement with the
underlying theoretical assumptions.

\subsection{Fit to $F_2$ data}

In this subsection we present results of the  fit to the low $x$
$F_2$ data. The structure function $F_2$ is given by a sum of
three contributions:

\beq\label{F2} F_2\,=\, \tilde F_2\,+\,\Delta F\,+\,F_2^{NSQ}\,,
 \eeq
where the first two terms are given by Eqs. (\ref{F2T}) and
(\ref{DF2}). These terms take into account only the gluon
contribution to $F_2$. In fact, gluons are related to singlet
quark distributions. The third term in (\ref{F2}) takes into
account contributions of non-singlet quark distributions:

\beq\label{NSQ} F_2^{NSQ}\,\,=\,\,\sum_{i=u,d}\,\, e_i^2\,
\,q_i^V\,. \eeq

In the present stage of our research we borrow the valence quark
distributions ($q_i^V$) from the LO CTEQ6 parametrization. It is
important to note, however, that these distributions decrease with
decreasing $x$ and are of practically no significance below
$x\simeq 10^{-3}$. In future  we plan to develop a fully self
consistent approach  without relying on any known parametrization.

\begin{figure}[htbp]
\begin{tabular}{c c}
(a) & (b) \\
 \epsfig{file=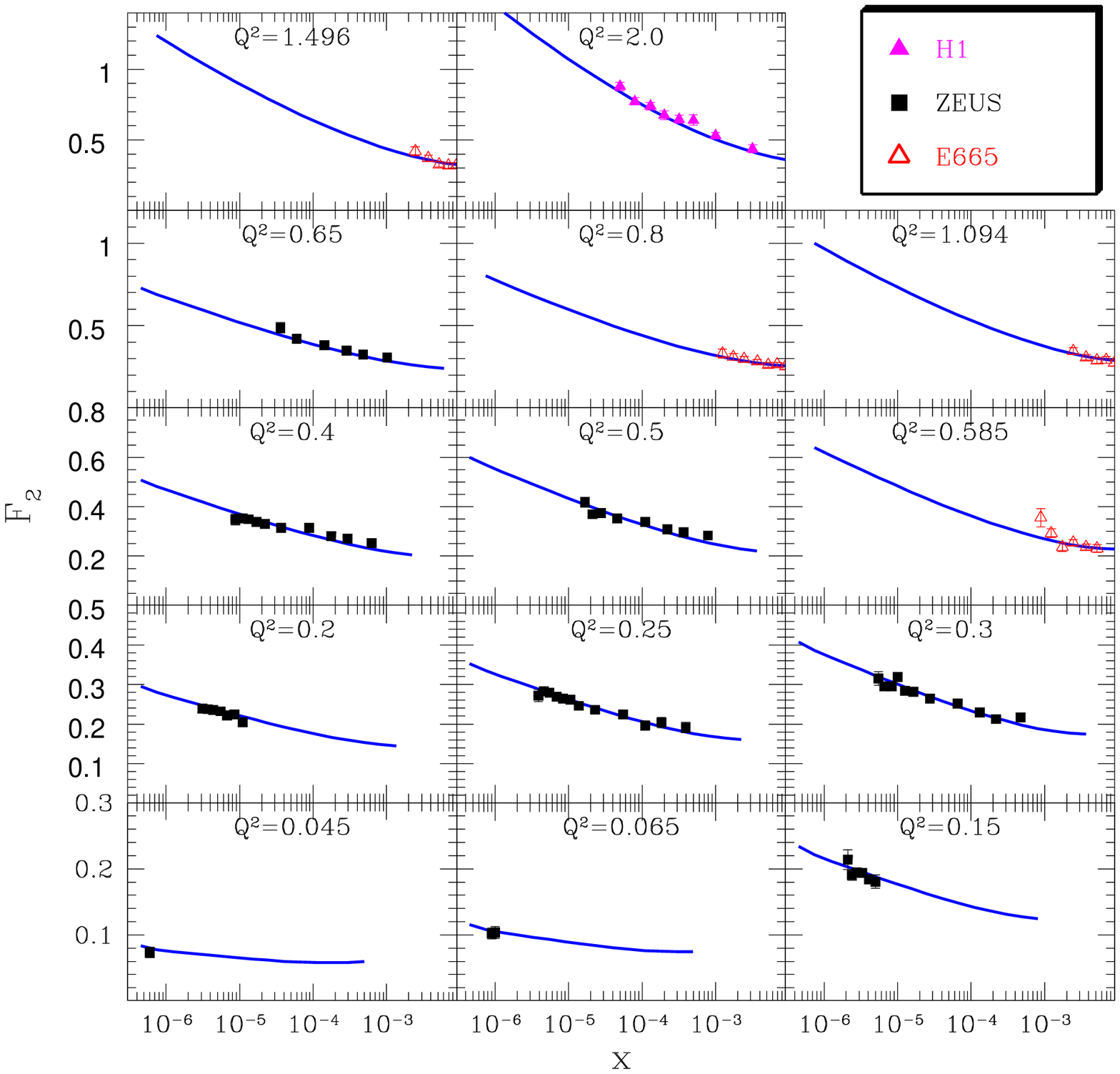,width=85mm, height=95mm}&
 \epsfig{file=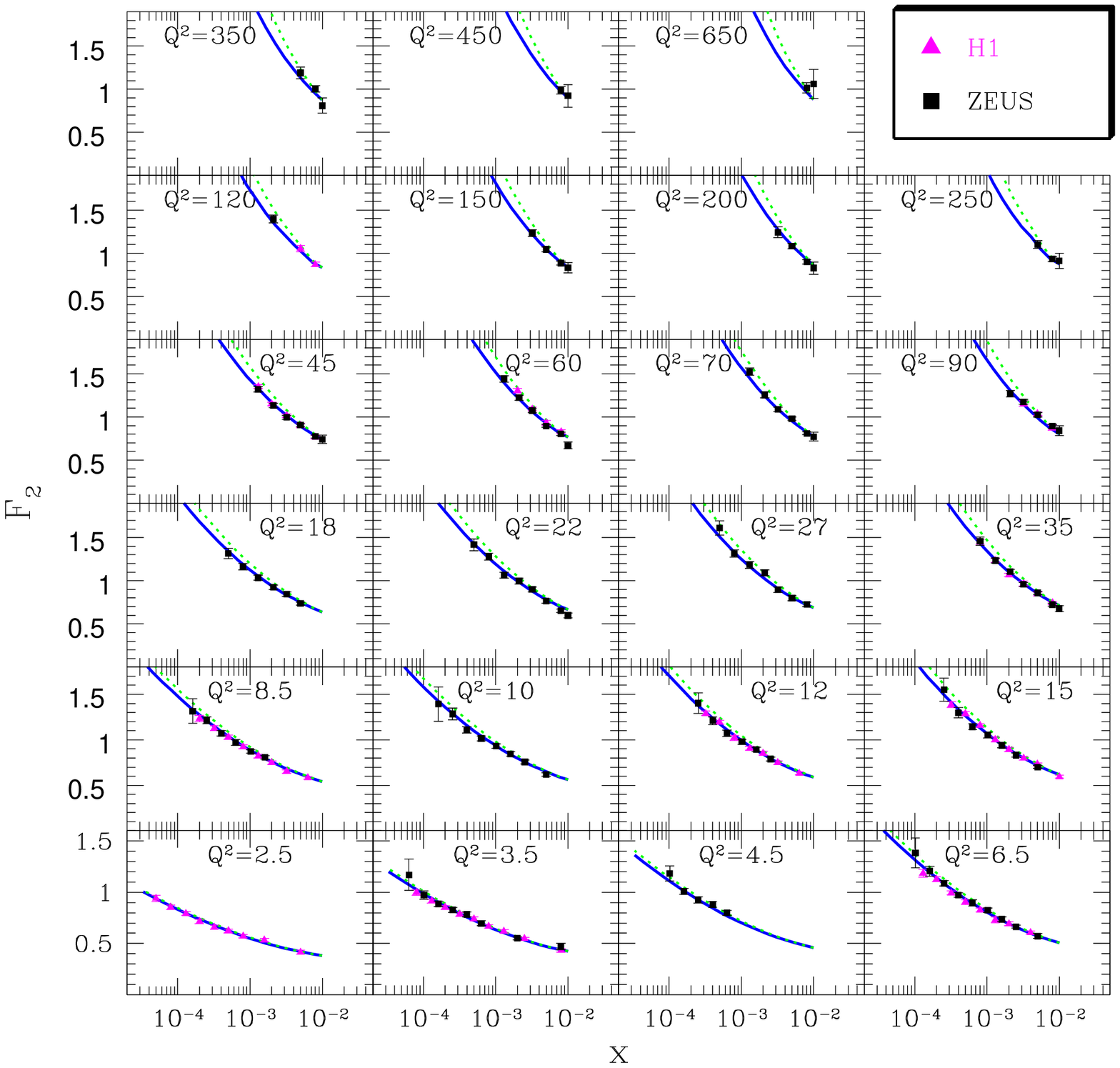,width=85mm, height=95mm}  \\
\end{tabular}
  \caption[]{\it Fit to the $F_2$ structure function. }
    \label{F2plot}
\end{figure}

Our central results are presented in Fig. \ref{F2plot}  for small
$Q^2$ (a) and  for large $Q^2$ (b). The solid  line is the
best fit obtained with resulting $\chi^2/df=1$. The dashed line
 is a result obtained without the DGLAP correction $\Delta
N$ ($\chi^2/df > 3$).

The quality of our fit is of the same level as of Ref. \cite{BGH}.
 A similar quality  fit was also obtained on a basis of a new
scaling saturation model of Ref. \cite{Munier}.

\subsection{$d F_2/d(\ln Q^2)$}

\begin{figure}[htbp]
\begin{tabular}{c c}
(a) & (b) \\
 \epsfig{file=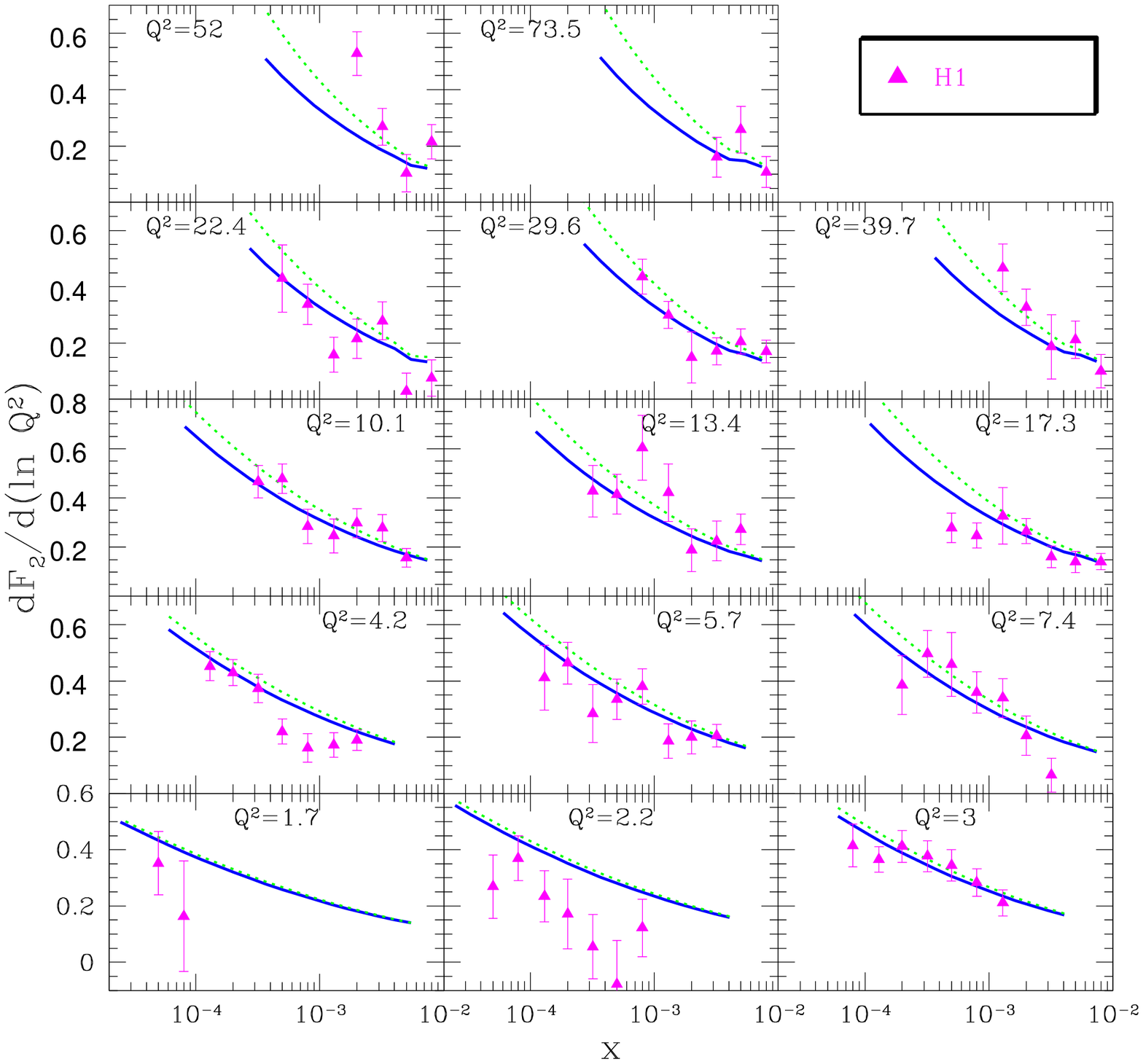,width=85mm, height=95mm}&
 \epsfig{file=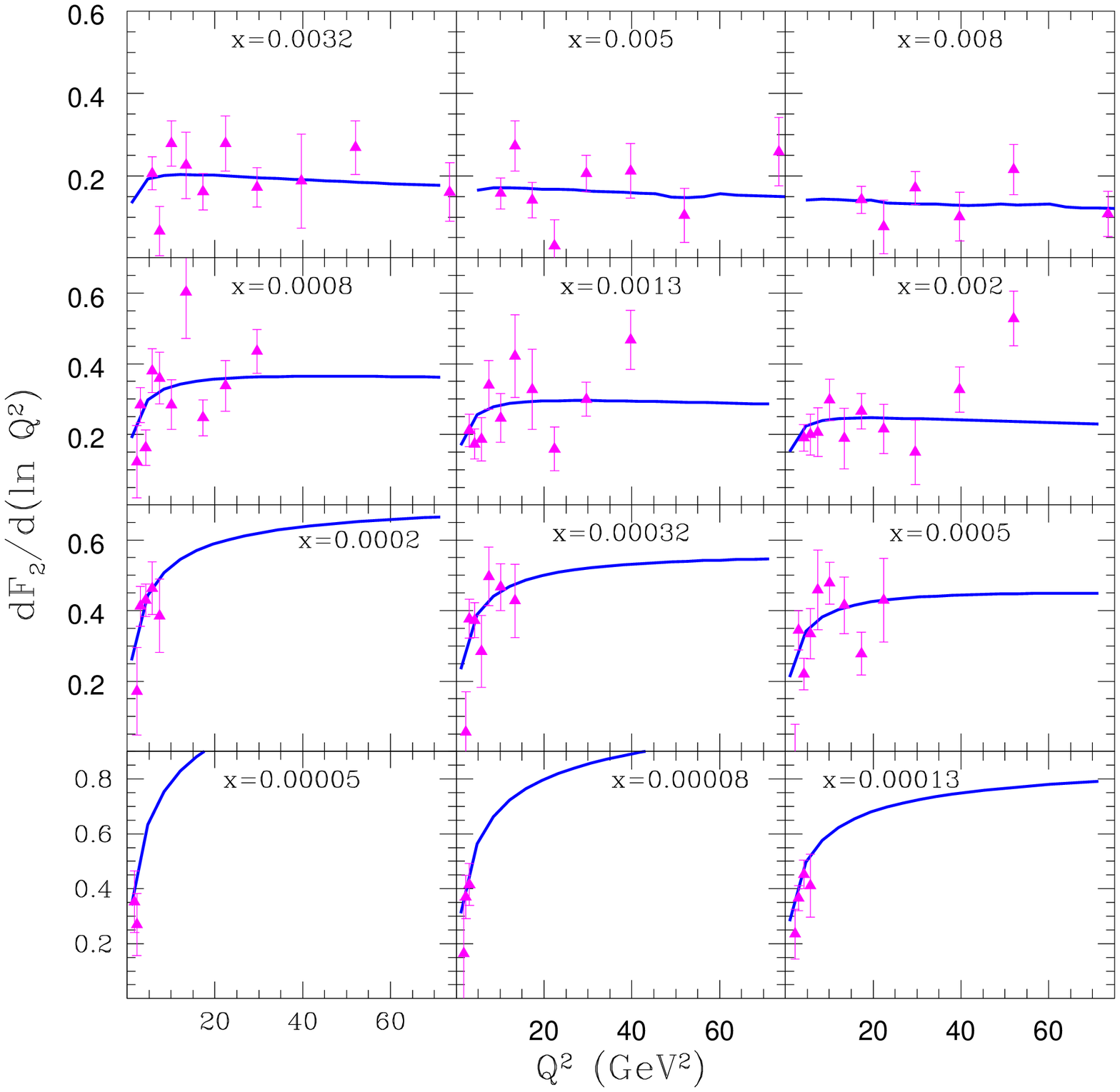,width=85mm, height=95mm}  \\
\end{tabular}
  \caption[]{\it The logarithmic derivative $\partial F_2/\partial \ln Q^2$. }
    \label{DF2DLQ}
\end{figure}

The logarithmic derivative of $F_2$ with respect to $\ln Q^2$ is
presented in Fig. \ref{DF2DLQ} at fixed $Q^2$ (a) and  at fixed
$x$ (b). Only comparison with H1 data \cite{H1F2} is shown though
similar ZEUS measurement exists as well \cite{ZEUSslop}. Note that
these experimental data were not take into account in the fitting
procedure.

\subsection{$d \ln F_2/d(\ln 1/x)$}

\begin{figure}[htbp]
\begin{tabular}{c c}
(a) & (b) \\
 \epsfig{file=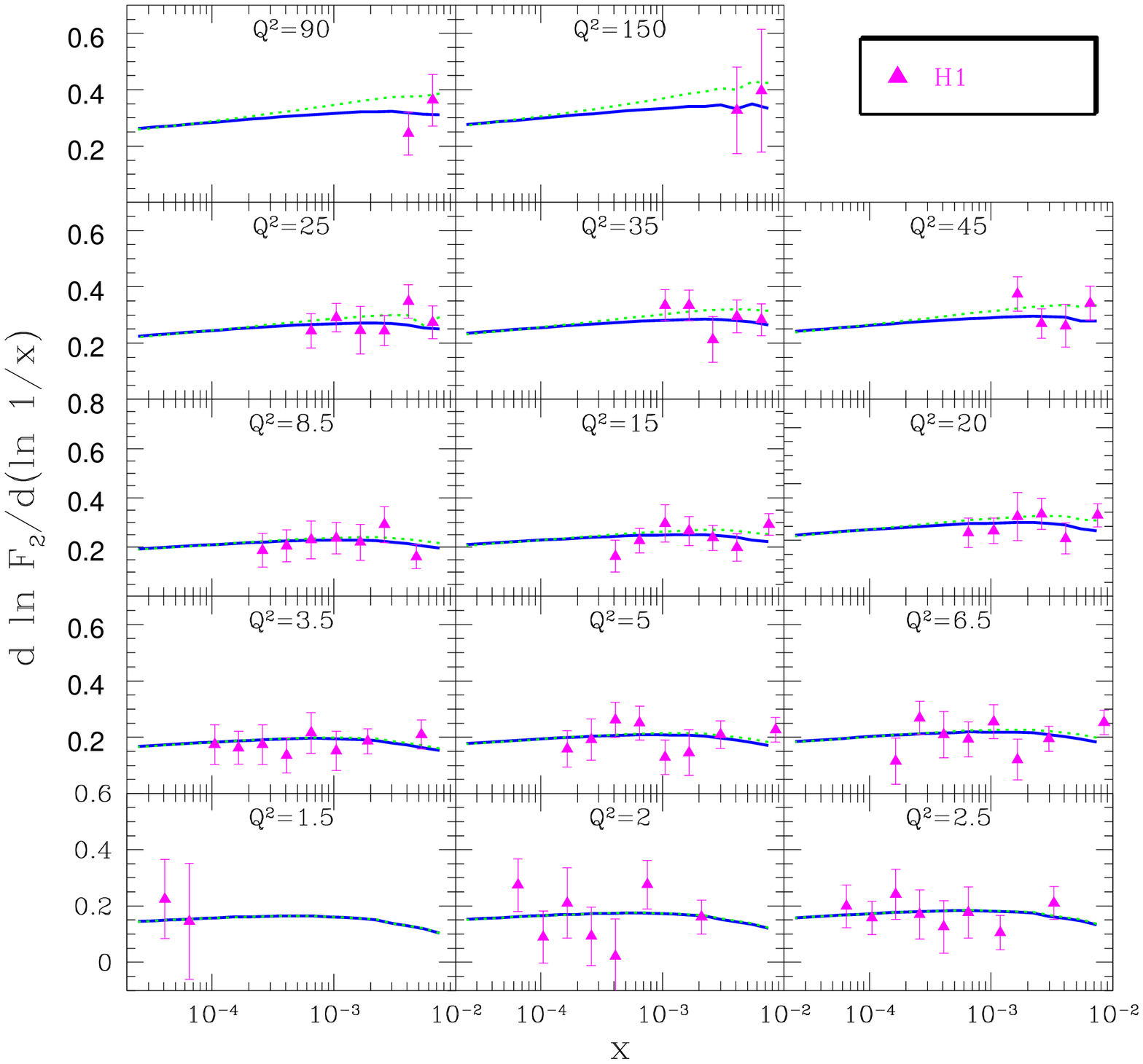,width=85mm, height=95mm}&
 \epsfig{file=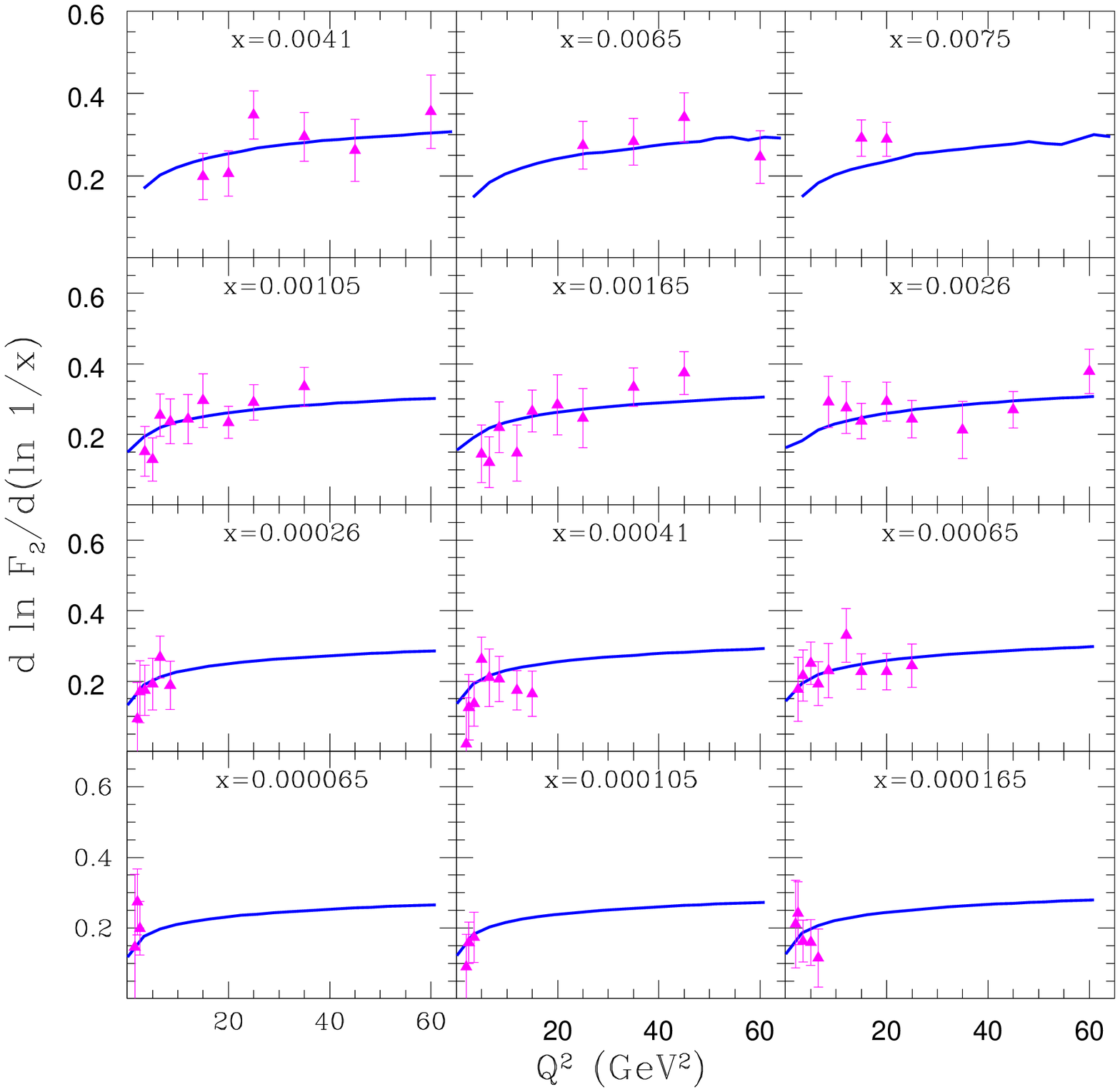,width=85mm, height=95mm}  \\
\end{tabular}
  \caption[]{\it The logarithmic derivative $\lambda\,=\,\partial \ln F_2/\partial \ln 1/x$. }
    \label{lamda}
\end{figure}

\begin{figure}[htbp]
\begin{minipage}{10.cm}
\epsfig{file=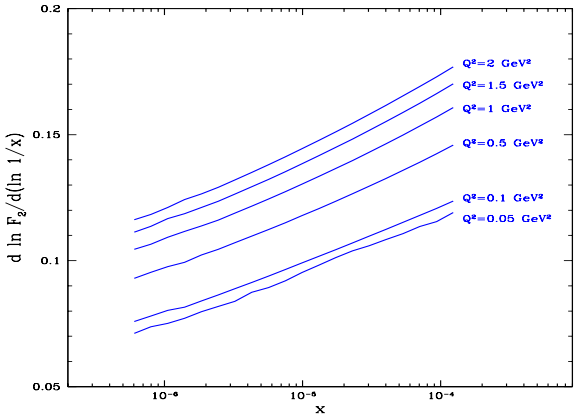,width=80mm, height=60mm}
\end{minipage}
\begin{minipage}{7 cm}
\caption{ \it The logarithmic derivative $\lambda\,=\,\partial \ln
F_2/\partial \ln 1/x$ plotted at low $Q^2$ and very low $x$.}
\label{lamdalow}
\end{minipage}
\end{figure}

In this subsection we present our computation of $\lambda\equiv \partial \ln F_2/\partial(\ln
1/x)$. A comparison with the H1 data \cite{H1lambda} is shown in
Fig. \ref{lamda} at fixed $Q^2$ (a) and at fixed $x$ (b).
Fig. \ref{lamdalow}  presents our prediction for $\lambda$ at very
low $x$ and small values of $Q^2$. At fixed $Q^2$, $\lambda $
decreases with decreasing $x$ tending to zero in agreement with
the unitarity constrain. At $Q^2$ well below $1\,GeV^2 $ and
$x\simeq 10^{-6}$, $\lambda \simeq 0.08 \pm 0.01$. This value of
$\lambda $ coincides with the "soft pomeron" intercept of the
Donnachie and Landshoff model (DL) \cite{DL}. It is important to
stress that the result is obtained on a basis of perturbative QCD.
The only nonperturbative input in our approach is the freezing of
$\as$ at large distances. In fact, it was conjectured in Ref.
\cite{CS} that the soft pomeron may appear in perturbative QCD due
to freezing of $\as$.

\subsection{Prediction for $F_L$ at HERA}

\begin{figure}[htbp]
\begin{minipage}{11.5cm}
\epsfig{file=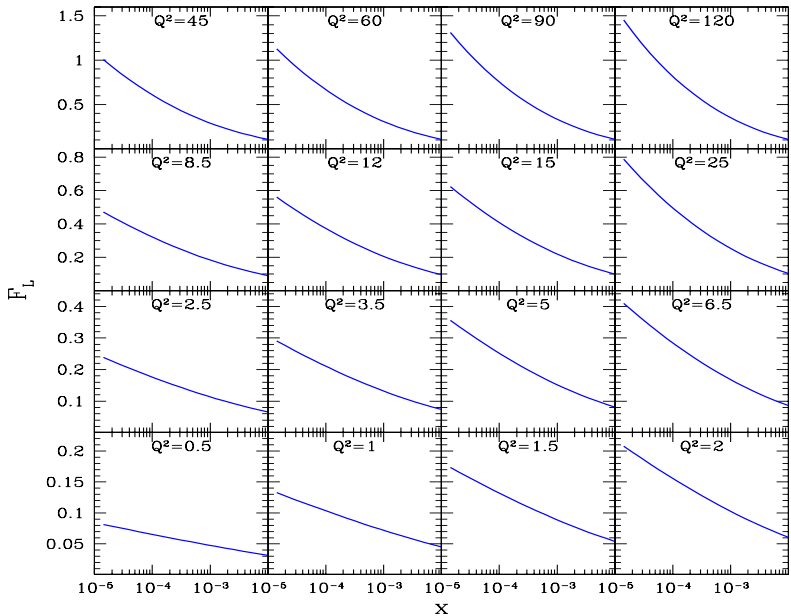,width=110mm, height=90mm}
\end{minipage}
\begin{minipage}{5.5 cm}
\caption{ \it Prediction for the $F_L$ structure function at the
HERA kinematics. The values of $Q^2$ are given in $GeV^2$.}
\label{FL}
\end{minipage}
\end{figure}

In this subsection we present our prediction for the $F_L$
structure function. The result is obtained on a basis of the
function $\tilde N$ (longitudinal part of $\tilde F_2$) (Fig.
\ref{FL}). The function $\Delta N$ is obtained within the leading
logarithmic approximation, and in this approximation it does not
contribute to $F_L$. Note that at relatively high values of
$x\simeq 10^{-2}\,-\,10^{-3}$ our prediction can be slightly
underestimated since the contribution of the valence quarks is
neglected.

\subsection{Predictions for LHC and THERA}

\subsubsection{Gluon density}

\begin{figure}[htbp]
\begin{tabular}{c c c}
\epsfig{file=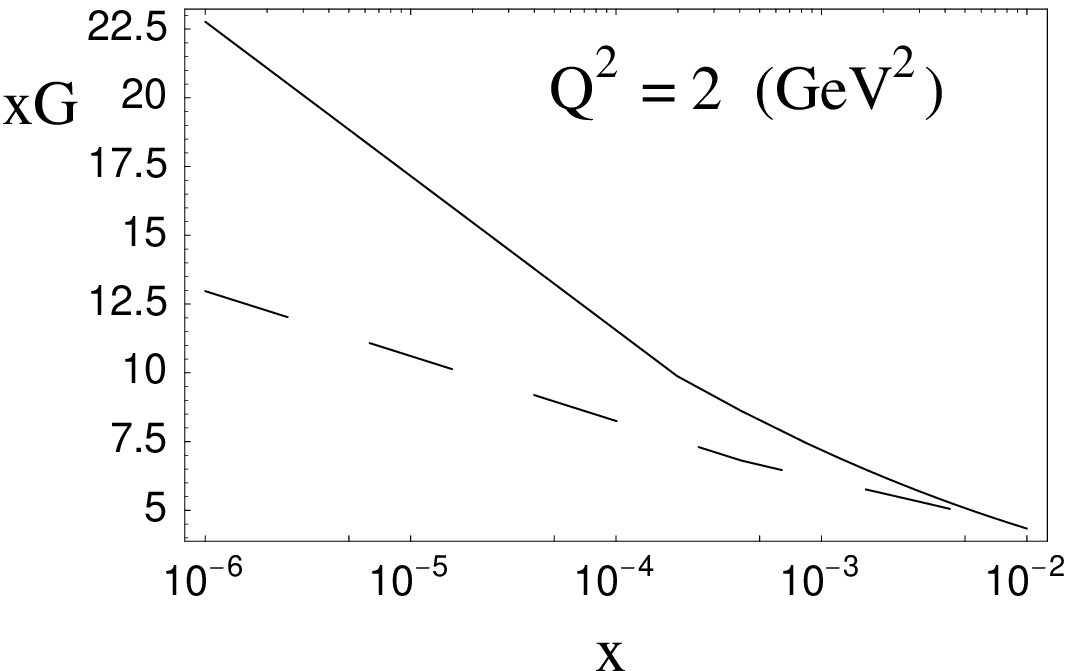,width=60mm, height=42mm}&
\epsfig{file=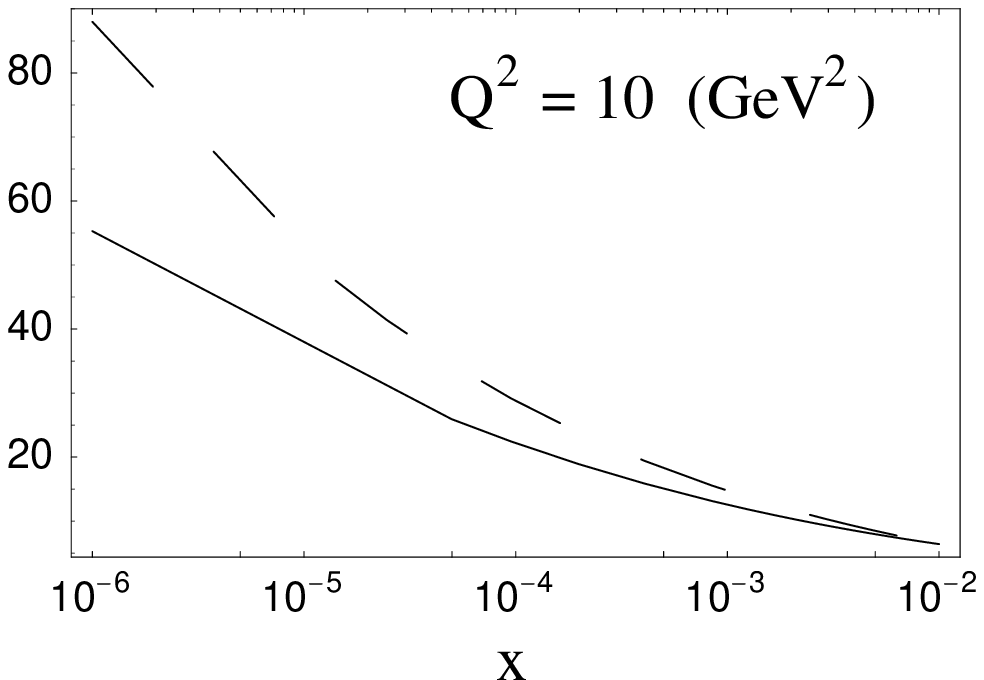,width=52mm, height=42mm}&
 \epsfig{file=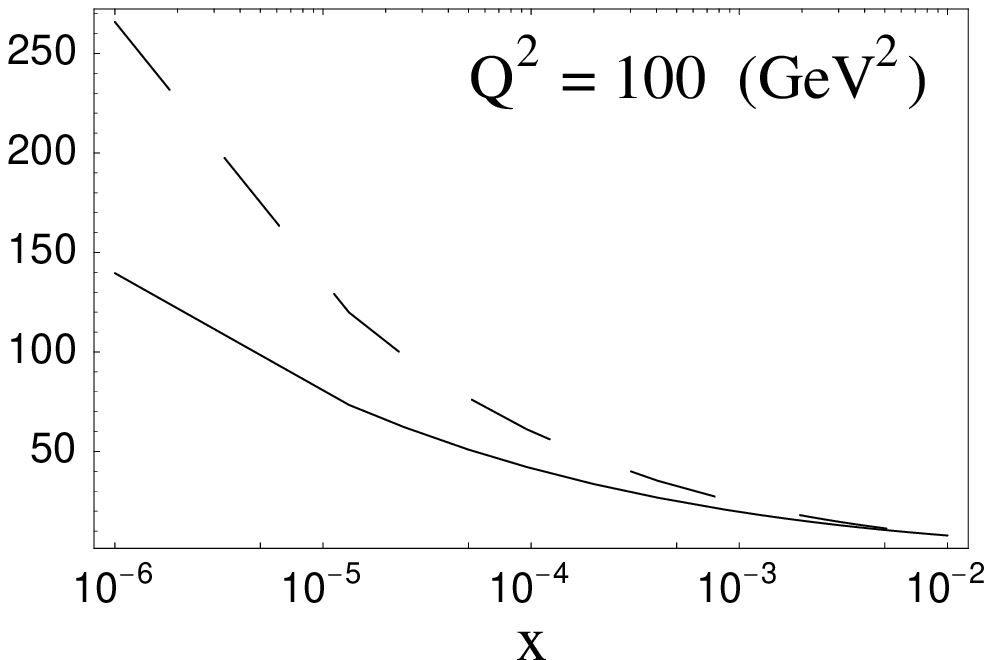,width=52mm, height=42mm}  \\
\end{tabular}
  \caption[]{\it The gluon density $xG$ is plotted versus $x$ at fixed $Q^2$.
   The solid line corresponds to Eq. (\ref{xG}) while the dashed line
   is for $xG^{DGLAP}$ (CTEQ6).}
    \label{xGfig}
\end{figure}

From the solutions obtained we can compute the gluon density. To
this goal we rely on the Mueller`s formula \cite{DOF3} which
relates the density to the elastic dipole-target amplitude:

\beq\label{xG} xG(x,Q^2)\,=\,\frac{4}{\pi^3}\,\int_x^1 \,\frac{d
x^\prime}{x^\prime}\,\int_{4/Q^2}^{r_{\perp\,0}}\,\frac{d
r_\perp^2}{r_\perp^4}\,\int\,d^2b\,\,2\,\,N(r_\perp,x^\prime)\,.
 \eeq
Practically we integrate in $x^\prime$ up to $x_0$ and then add
$xG^{DGLAP}(x_0)$. Fig. \ref{xGfig} presents a comparison between
the gluon density obtained from Eq. (\ref{xG}) and $xG^{CTEQ}$. At
very low $x$ a significant damping of the density can be observed
compared to the DGLAP predictions.

\subsubsection{$F_2$}

\begin{figure}[htbp]
\begin{minipage}{11.5cm}
\epsfig{file=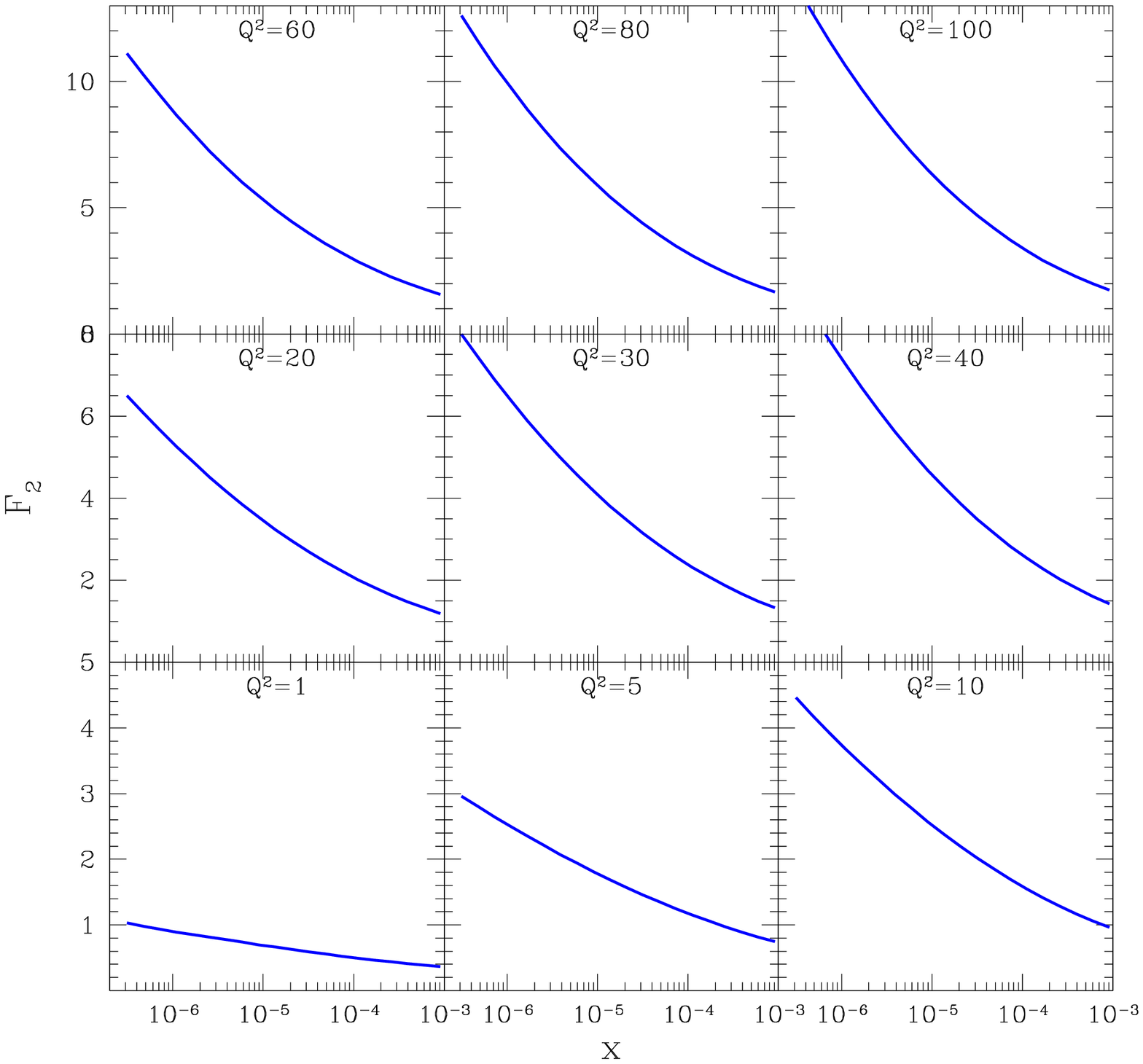,width=110mm, height=90mm}
\end{minipage}
\begin{minipage}{5.5 cm}
\caption{ \it Prediction for the $F_2$ structure function at the
LHC kinematics. The values of $Q^2$ are given in $GeV^2$.}
\label{F2LHC}
\end{minipage}
\end{figure}

The obtained model allows extrapolation of the parton
distributions to very high energies. Fig. \ref{F2LHC}  presents
our predictions for THERA and LHC kinematics.

\section{Discussion}

\subsection {The transverse hadron size $R^2$.}

As was pointed out above the optimal fit is achieved for
$R^2=3.1\,(\rm GeV^{-2})$. Such a low value requires understanding
and below we present several explanatory  arguments.

\begin{figure}[htbp]
\begin{minipage}{9.0cm}
\epsfig{file=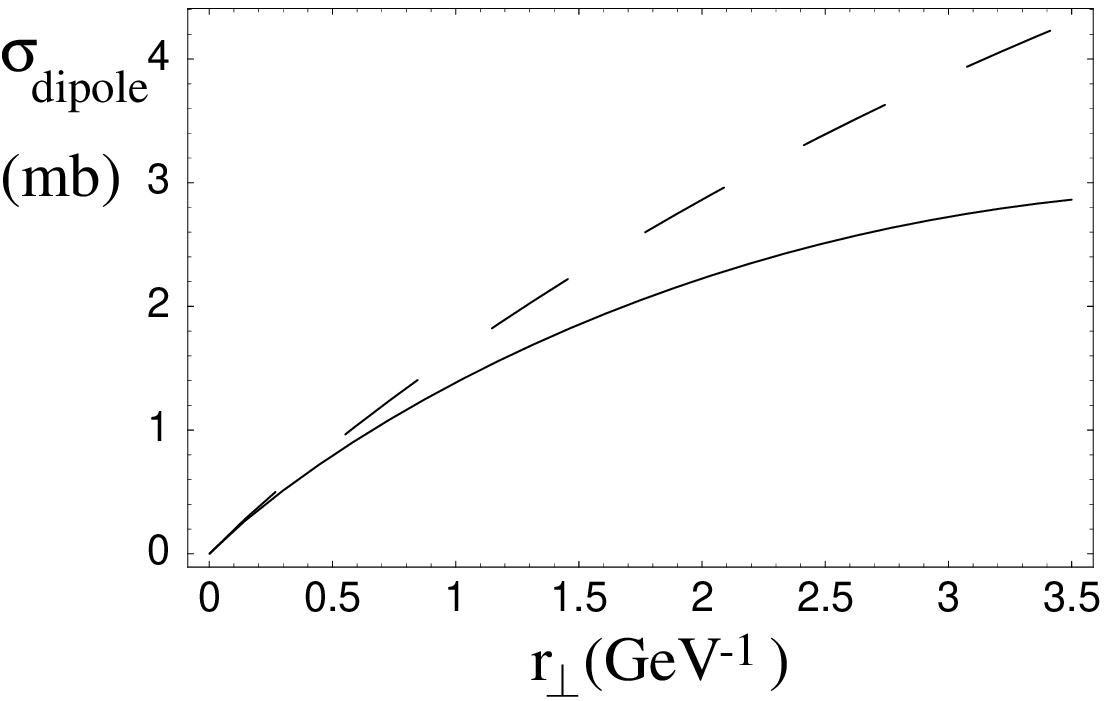,width=60mm, height=40mm}
\end{minipage}
\begin{minipage}{7.0 cm}
\caption{ \it The dipole cross section is plotted versus $r_\perp$
at $x= 10^{-2}$ for \eq{XSFIN} (upper curve) and for \eq{XSDIP}
(lower curve). $\as=0.2$ for this plot.} \label{xsdipplot}
\end{minipage}
\end{figure}

\begin{enumerate}
\item \quad First of all, the Glauber - Mueller formula of
\eq{Glauber} can be used for a proton target with great
reservations because  contrary to the nuclear case  large
inelastic diffraction is present. This process not only has a
considerable cross section but also a quite different impact
parameter dependence  corresponding  to a small value of the
radius.  The effect of two different radii is in agreement with
the HERA data on elastic and  inelastic $J/\Psi$ photo-production
\cite{HERAPSI}. In the simple additive quark model (AQM) there are
two kinds of processes: rescattering of a dipole off one quark and
rescattering due to interaction with two or even three constituent
quarks. The admixture of the inelastic diffractive processes can
be taken into account (see Ref. \cite{GLM2R} ) by effective
decreasing of $R^2$ in \eq{Glauber} from 10\,$(\rm GeV^{-2})$ to
5\,$(\rm GeV^{-2})$.

\item \quad Second, in \eq{Glauber} we used the Gaussian
parametrization for $b$-dependence. On the other hand, the data on
$J/\Psi$ production requires assuming the profile function of the
form \cite{PSIGLM}: \beq \label{DIPOLES}
S(b)=\frac{1}{\pi\,R^2}\frac{\sqrt{8}b}{R}K_1(\frac{\sqrt{8}b}{R})
\eeq which corresponds to the power - like (dipole) form factor in
momentum transfer representation: \beq \label{MTRFF}
F_{dipole}(t)=\frac{1}{(1-\frac{1}{8}R^2\,t)^2}\,. \eeq
$F_{dipole}(t)$ describe a system with the same radius $R$ as the
Gaussian form factor. In t-representation the latter looks as \beq
\label{EXPFF}
 F_{exp}(t)=e^{\frac{1}{4}R^2\,t}\,\,.
\eeq

 Practically,
the solution to the non-linear BK evolution equation is obtained
at $b=0$. Note that $S_{dipole}(b =0) = 2 \,S_{Gaussian} (b=0)$ and
 this difference  can be interpreted as an effective decrease in
the value of $R^2$. In fact, a relatively good fit to the low $x$
data can be obtained with the dipole profile function at
$R^2\simeq 4.5\, (\rm GeV^{-2})$.

\item \quad In \eq{Glauber} we use the following expression for the
dipole - proton cross section: \beq \label{XSDIP} \sigma_{dipole}
\,=\,\frac{\as\, \pi^2 \,r^2_\perp}{ N_c}
xG(x,\frac{4}{r^2_\perp})\,\,, \eeq However, the expression
\eq{XSDIP} is  an approximation valid for small values of the
anomalous dimension $\gamma$ only. The correct expression for the cross section was
obtained in Ref. \cite{SCPSIGLM}:

\beq \label{dcs}
\sigma_{dipole}(r_\perp,x)\,=\,\frac{16\,C_F}{N_c^2-1}\,\pi^2\,\int\phi(x,l^2)\,\,(1-e^{i\,l\,r_\perp})\,
\frac{\as(l^2)}{2\pi}\,\frac{d^2l}{l^2}
\eeq
with $\phi\equiv \partial xG(x,l^2)/\partial l^2$ being an
unintegrated gluon density \cite{BFKL}. The gluon density
$xG\equiv xG^{DGLAP}$ is a solution of the DGLAP equation
\beq \label{xgdglap}
xG(x,l^2)\,=\,\frac{1}{2\,\pi\, i}\,\int_C
d\omega\,x^{-\omega}\,g(\omega)\, e^{\gamma(\omega)\,\ln l^2}\,.
\eeq
Substituting Eq. (\ref{xgdglap}) into Eq. (\ref{dcs}) and performing
the $l$-integration we obtain:

\beq \label{XSDIPC} \sigma_{dipole} \,=\,\frac{4\,\as\, \pi^2 }{
N_c}\,\int_{C}\,\,\frac{d \omega}{
2\,\pi\,i}\,\,g(\omega)\,\,e^{\omega\,\ln(1/x)}\,\frac{1}{1 -
\gamma(\omega)}\,\,\,\left(\,\frac{r^2_\perp}{4}
\right)^{1\,-\,\gamma(\omega)}\,\,\frac{\Gamma(
1\,+\,\gamma(\omega))}{\Gamma( 2\,-\,\gamma(\omega))}\,\,. \eeq

 It turns out that \eq{XSDIPC} can be approximately rewritten in the
very compact form: \beq \label{XSFIN} \sigma_{dipole}
\,=\,\frac{4\,\as\, \pi^2 }{ N_c}\,\int^{\frac{r^2_\perp}{4}}\,\,d
\,r'^2_\perp\,\,xG(x,\frac{1}{r'^2_\perp})\,\,. \eeq

\eq{XSFIN} and \eq{XSDIP} are quite different (see \fig{xsdipplot})
which again can be taken into account by reducing the value of
$R^2$.
\end{enumerate}

\subsection{Geometrical scaling and saturation scale}

\begin{figure}[htbp]
\begin{minipage}{9.0cm}
\epsfig{file=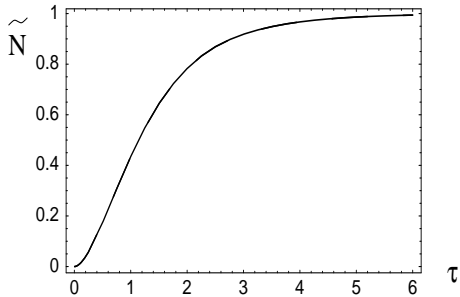,width=60mm, height=40mm}
\end{minipage}
\begin{minipage}{7.0 cm}
\caption{ \it Geometrical scaling. $\tilde N$  versus $\tau=r_\perp\, Q_s(x)$. } \label{scaling}
\end{minipage}
\end{figure}

We  briefly discuss the issue of the geometrical scaling displayed
by the function $\tilde N$. Namely $\tilde N =\tilde N (\tau)$
with $\tau \equiv r_\perp Q_s(x)$. The phenomena of geometrical
scaling for a solution of the BK equation was studied analytically
in Ref. \cite{LT} and established numerically in Ref.
\cite{Braun2,me1,GMS}. Recall that in the extrapolation of the
initial conditions to the very long distances we relied on the
scaling property. The function $\tilde N$ is displayed as a
function of $\tau$ in Fig. \ref{scaling}.

Three comments are in order.
\begin{itemize}
\item Fig. \ref{scaling} is obtained assuming $Q_s(x=10^{-3}) = 1\, (\rm
  GeV)$.
\item Within 10\% accuracy the scaling holds for $\tau \ge 1$. For
       smaller $\tau$ there is a noticeable scaling violation depending
       on the value of $x$. In fact, more significant scaling
       violation is found in perturbative region compared to the
       results of Ref. \cite{me1}. This discrepancy is likely to be due
       to the difference between $\as$: in Ref. \cite{me1}, a
       constant value of $\as=0.25$ was used while the present work is
       done with a running $\as$. It was argued in Refs. \cite{KS,IIM} that
       the running of $\alpha_s$ provides an important source for
      scaling breakdown,  when penetrating the region of perturbative QCD.
\item
  The $x$-dependence of the saturation scale can be investigated using
  the scaling property. Parameterizing $Q_s \sim x^{-q}$ we find
  $q=0.18\pm 0.02$. This value is about half the size of
  previous estimates of Refs. \cite{LL,me1}. The latter were obtained
  with the constant $\as=0.25$ and the decrease of $q$ is doubtless
   due to running of $\alpha_s$. Indeed, in Ref. \cite{LGLM} we showed that in the
   case of running $\as$, the saturation scale grows much slower
   than the fixed constant case. It is certainly interesting to investigate
   the dependence of $q$ on $\alpha_s$.
\end{itemize}

\subsection{Comparison with the GBW model}

\begin{figure}[htbp]
\begin{tabular}{c c c}
\epsfig{file=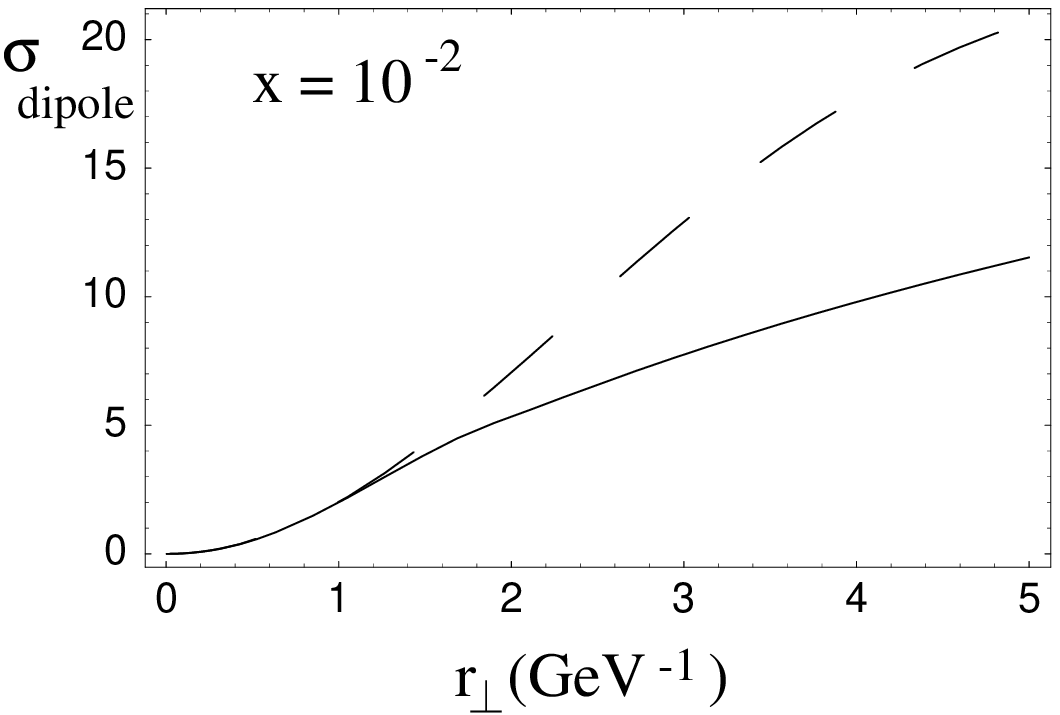,width=54mm, height=42mm}&
\epsfig{file=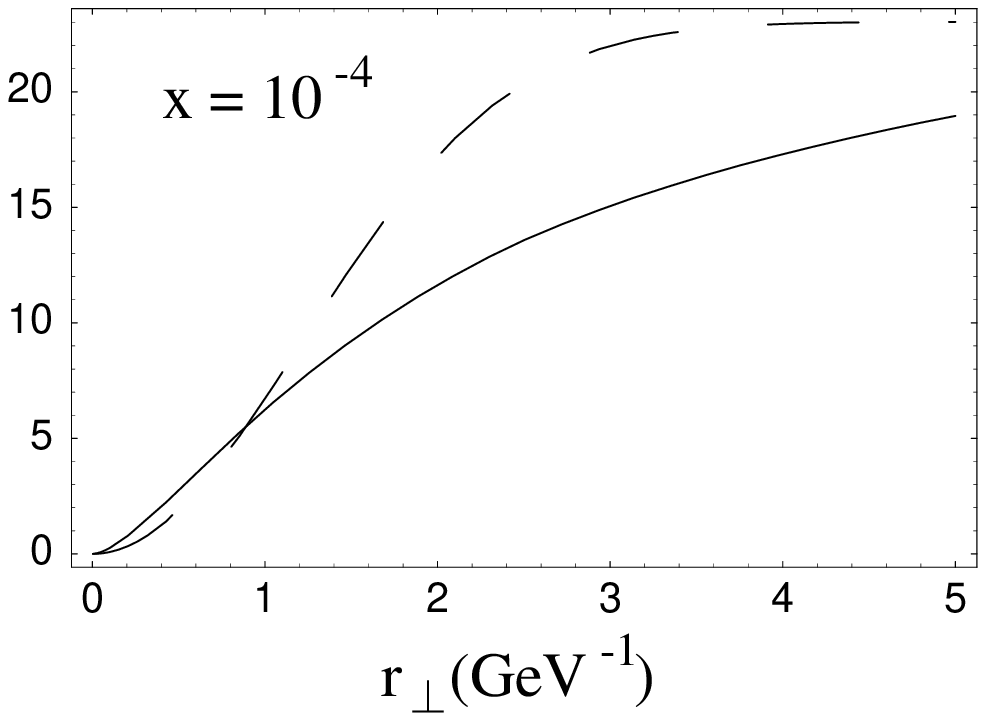,width=54mm, height=42mm}&
 \epsfig{file=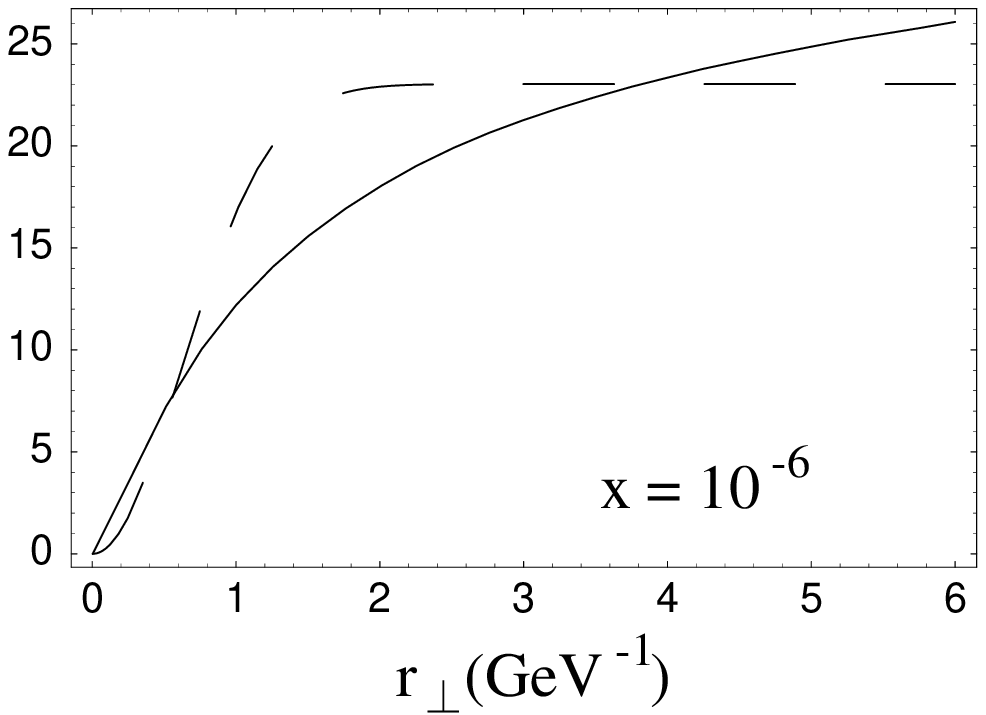,width=54mm, height=42mm}  \\
\end{tabular}
  \caption[]{\it The comparison between $\sigma_{dipole}$
    (\ref{sidipol1}) (solid curve) and the dipole cross section of the
    GBW model (dashed curve).}
    \label{gbw}
\end{figure}

It was mentioned in the Introduction that the solution to the BK
equation ($\tilde N$) in describing the low $x$ data plays the same
role as the GBW original saturation model. Consequently, it of
interest to compare these two models. Fig. \ref{gbw} shows the dipole
cross section  (\ref{sidipol1}) plotted together with the one of the
GBW model. Note that due to the impact parameter integration the
dipole cross section  (\ref{sidipol1}) grows with decreasing $x$
(logarithmically) while the GBW model reaches a saturation value.

As a function of $r_\perp$ the behavior of the curves in Fig.
\ref{gbw} is quite different. This is a numerical coincidence that
after the $r_\perp$ integration these dipole cross sections
(improved by the DGLAP corrections) lead to a good description of
the very same data.

\subsection{Shortcomings of our approach}

We would like to list several shortcomings of our approach and
indicate future steps for their elimination.

\begin{itemize}

\item One of the theoretical difficulties is in the fact that
$\Delta N \,\,\rightarrow\,\,const(x) < \,0 $ at low $x$ and fixed
$b$. In spite of the fact that this limit does not contradict the
unitarity constraints, it looks very unnatural that the dipole
amplitude does not reach the maxim possible value $\tilde N +
\Delta N = 1$. Indeed, this fact is an artifact of our
approximation, namely, of the oversimplified form of the
non-linear term in \eq{EQN} in which $ 1/z$ should be also
replaced by the full kernel $P_{g \rightarrow g}(z)$. In  future
we plan to treat this problem. Here, we want to recall that after
integration over $b$, $\int d^2 b\, \Delta N$ becomes much smaller
than $\int \,d^2 b\,  \tilde N$ due to the logarithmical grows of
the latter as a function of $x$.

\item Our results are based on the CTEQ parametrization, which
enters our calculations through gluon distribution at $x\ge
10^{-2}$ and valent quark distributions.

We attempted to switch to another parametrization (GRV98
\cite{GRV}) but failed to reproduce a very good fit ($\chi^2/df
\simeq 2.3$). The main reason for this failure is because at
$x_0=10^{-2}$ and  very short distances the gluon of GRV is
smaller than the CTEQ gluon, by about 10\%. This difference cannot
be practically eliminated by adjusting of our fitting parameter
$R$. Yet, the treatment of the dipole cross section in the from
presented by Eq. (\ref{XSFIN}) is  likely to improve the
situation.

At present, we have to conclude that our current results are
parametrization dependent. This requires  us to reconsider the
problem by producing our own DGLAP fit of high $x$ data which
would also include the quark distributions.

\item
 One of the central uncertainities of our approach is in the
impact parameter dependence of the function $\tilde N$. The ansatz
(\ref{Nb}) is certainly not fully correct though it preserves the
main properties of the $b$-dependence. In our approach, the
uncertainty due to this ansatz was partly  hidden in the fitting
of the effective target size $R^2$. In order to eliminate this
problem it is highly desirable to solve Eq. (\ref{EQ}) including
the full $b$-dependence of the solution.

\end{itemize}

\section{Summary}

A new approach to DIS based on summation of high twist
contributions in the leading $\ln x$ approximation is developed.
The first step implies solution of the Balitsky-Kovchegov
nonlinear evolution equation. Secondly, a linear evolution
equation for the correcting function which incorporates the
correct DGLAP kernel in the leading $\ln Q^2$ approximation is
derived. It is important to stress that both the equations are
based on QCD and derived in several approximations.

The BK equation (\ref{EQ}) is solved numerically by the method of
evolution. The solution leads to a saturation of the function
$\tilde N$ at large distances. However, the dipole cross section
obtained is not saturated as a function of $x$.  Due to the $b$
integration   it grows logarithmically with decreasing $x$ in a
contrast to the GBW saturation model.

The DGLAP correcting function $\Delta N$ was found as a solution
of Eq. (\ref{DN}). In agreement with the analytical estimates this
function contributes at moderate values of $x$ and provides a
correction to the main contribution due to $\tilde N$.

As a main goal of this work, the low $x$ $F_2$ data is fitted in
the whole kinematic region both for small and large photon
virtualities $Q^2$. The resulting $\chi^2/df=1$. In order to
achieve this result practically only one fitting parameter is
used. This fit determined the optimal model which later is applied
to compute logarithmic derivatives of $F_2$. Several predictions
for the THERA and LHC are presented.

Analyzing $\lambda\equiv \partial \ln F_2/ \partial (\ln 1/x)$ at
very low $x$ and small photon virtualities we found $\lambda
\simeq 0.08$ which coincides with the "soft pomeron" intercept of
the DL model. It is important to stress that this soft pomeron
occurs as an effective result of multiple hard (BFKL) pomeron
rescattering. Except freezing of $\as$, no soft physics is
introduced in our approach.

The obtained model opens a possibility to address many questions
in high energy phenomenology. At present we are working on DIS
 off  nuclei as well as  more  exclusive
processes such as of $J/\psi$ production.

{\bf Acknowledgements:} We would like to thank the DESY Theory
Group for their hospitality and creative atmosphere during several
stages of this work.

We wish to thank Krzysztof Golec-Biernat,  Dima Kharzeev, Larry
McLerran, Eran Naftali, and Guenter Grindhammer as well as all
participants of Lund meeting of ``Small x collaboration" for very
fruitful discussions.

This research was supported in part by the BSF grant $\#$
9800276, by the GIF grant $\#$ I-620-22.14/1999
  and by
Israeli Science Foundation, founded by the Israeli Academy of Science
and Humanities.

\end{document}